\documentclass[%
 reprint,
superscriptaddress,
showkeys,
preprintnumbers,
amsmath,
amssymb,
aps,
prd,
floatfix,
]{revtex4-2}

\usepackage{caption}
\usepackage{graphicx}
\usepackage{dcolumn}
\usepackage{bm}
\usepackage[compatibility=false]{caption}
\usepackage[caption=false]{subfig}
\usepackage{hyperref}

\usepackage{ulem}
\usepackage[dvipsnames]{xcolor}

\usepackage{xcolor}
\definecolor{darkblue}{rgb}{0,0,0.5} 

\hypersetup{
    colorlinks=true,
    linkcolor=darkblue,
    citecolor=darkblue,
    urlcolor=darkblue
}
\usepackage{booktabs}

\begin{document}


\title{Dark energy from neutrino interactions in  Unimodular Gravity}

\author{A. Gil-Ocaranza }
\email{victor.gil@academicos.udg.mx}
\affiliation{Departamento de Física, Centro de Investigación y de Estudios Avanzados del Instituto
Politécnico Nacional
Apartado Postal 14-740, 07000, Ciudad de México, México.}

\author{Josue De-Santiago}
\email{Josue.desantiago@cinvestav.mx}
\affiliation{Departamento de Física, Centro de Investigación y de Estudios Avanzados del Instituto
Politécnico Nacional
Apartado Postal 14-740, 07000, Ciudad de México, México.}
\affiliation{Secretaría de Ciencia, Humanidades, Tecnología e Innovación, Av. Insurgentes Sur 1582,
03940, Ciudad de M\'exico, M\'exico}
\author{Mauricio Lopez-Hernandez}
\email{mauricio.lopez@cinvestav.mx}
\affiliation{Departamento de Física, Centro de Investigación y de Estudios Avanzados del Instituto
Politécnico Nacional
Apartado Postal 14-740, 07000, Ciudad de México, México.}

\author{Jorge L.~Cervantes-Cota}
\email{jorge.cervantes@inin.gob.mx}
\affiliation{%
Departamento de F\'isica, Instituto Nacional de Investigaciones Nucleares,
Apartado Postal 18-1027, Col. Escand\'on, Ciudad de M\'exico, 11801, M\'exico.}

\date{\today}

\begin{abstract}
We investigate a dark energy scenario generated by neutrino interactions mediated by a light 
scalar field, in which finite-temperature corrections induce an effective neutrino mass that 
evolves with the thermal history of the Universe. Within the framework of Unimodular Gravity, 
these interactions give rise to a non-conservation current, leading to dynamical dark energy. We study one- and two-neutrino 
realizations of the  model. In the one-neutrino case, the dark energy density evolves 
monotonically, whereas in the two-neutrino scenario it can reach a maximum at intermediate 
redshifts before decreasing at late times.  Using late time cosmological datasets, we constrain the effective interaction 
strength for lightest-neutrino masses in  the range $0.05 \,{\rm meV}\le m_1 \le 1 \,{\rm meV}$. We find preferred interaction scales of 
order $G_s\sim10^{12} \, {\rm eV}^{-2}$ with a significance of $2 \sigma$, with the inferred coupling decreasing as the assumed 
neutrino mass increases. Assuming neutrino couplings of order unity, this $G_s$ value 
corresponds to an ultralight mediator with mass $m_\phi\sim10^{-6} \, {\rm eV}$. We further 
assess the impact of Planck distance-prior,
finding a noticeable reduction in parameter degeneracies and a reconstructed dark energy evolution closer to that of a cosmological 
constant. Our results show that neutrino interactions can generate both monotonic and 
non-monotonic dark energy evolutions while remaining compatible with current cosmological 
observations. The inferred interaction strengths remain consistent with non-zero values  for 
part of the explored neutrino-mass range, supporting neutrino-induced dark energy dynamics 
as a viable phenomenological extension of $\Lambda$CDM at the background level.
\end{abstract}


\maketitle

\tableofcontents

\begin{section}{Introduction}

The recent cosmological analysis made by the DESI collaboration implies the 
possibility that dark energy may be an evolving function, instead of the constant 
$\Lambda$. This evidence comes from the joint analysis of the newest BAO DESI dataset \cite{DESI:2024mwx, DESI:2025zgx}, in combination with Planck CMB \cite{Planck:2019nip} 
and supernovae type Ia  (SNe) data, including 
Pantheon+ \cite{Scolnic:2021amr,Brout:2022vxf}, Union3 \cite{Rubin:2023jdq}, and DES year 5 (DESY5) 
\cite{DES:2024jxu}. This possibility was also backed by other DESI dark energy analyses \cite{DESI:2024aqx,DESI:2024kob,DESI:2025fii} and the DESI  BAO+fullshape analysis 
\cite{DESI:2024hhd}, among other results from outside the DESI collaboration, although a debate on 
this issue is still ongoing. These results indicate that, although the best-fit parameter inference is 
consistent with the $\Lambda$CDM model, the datasets favor an evolving behavior that could be compatible with the 
$w_0w_a$ (CPL) dark energy parameterization or with other more complex functions 
\cite{DESI:2024aqx,DESI:2024kob,DESI:2025fii}.

On the other hand, the aforementioned analyses also appear to favor an evolving dark energy component when the sum of neutrino masses ($\Sigma m_\nu$) is treated as a free parameter. Within $\Lambda$CDM, 
the best fitted value for $\Sigma m_\nu$ results in a close-to-zero value, $\Sigma m_\nu < 0.0642$ 
eV ($95 \%$ C.L.) \cite{DESI:2025zgx}, compromising the known outcome of ground-based neutrino mass 
experiments in which $\Sigma m_\nu > 0.059$ eV for normal mass ordering and $\Sigma m_\nu > 0.1$ eV 
for inverted mass ordering \cite{Esteban:2024eli}; more recent results from SPT-3G D1+DESI and 
CMB-SPA+DESI combinations appear to rule out the normal and inverted hierarchies at $97.9\%$ and 
$99.9 \%$ confidence, respectively \cite{SPT-3G:2025bzu}.  Related to these results, the posterior probability in the $\Lambda$CDM model peaks in the region of negative $\Sigma m_\nu$, indicating a tension between the preferred cosmological fit and the physical requirement of positive neutrino masses. The $w_0 w_a$CDM model, in contrast, seems to 
alleviate this tension, but it depends on the datasets considered. For instance, when SNe data are 
included in the analysis again a negative mass is favored \cite{Elbers:2025vlz}. The inclusion of 
curvature again does not solve the problem \cite{Pulido-Hernandez:2026hcs}; the only means to solve this tension appears to be increasing the reionization optical depth parameter $\tau$ \cite{Sailer:2025lxj}, but to values that are not compatible with the low-$\ell$ CMB results.

From the particle physics perspective, neutrinos present several open questions, including oscillation anomalies in the terrestrial experiments LSND and MiniBooNE \cite{LSND2001,MiniBoone2018,Palomares2005}. Neutrino nonstandard interactions offer a possible resolution to these anomalies and, in turn, carry important cosmological implications. Numerous works have studied the cosmological consequences of such interactions \cite{Bell2006,Archidiacono2014,CyrRacine2014,Lancaster_2017,Park2019,Escudero:2019gvw,Kreisch2020,knox2020hubble,Choudhury2021,Venzor:2023aka,Oldengott2015,Lancaster2017,Oldengott2017,Beacom2004,Hannestad2005,Chacko2020,EscuderoRelaxing,Esteban2021,Barenboim2021,abellan2021, FrancoAbellan:2026ori,Noriega:2025ulc,Perez-Castro:2026muj}; they found, among other effects, that the interactions can relax the cosmological bounds to the neutrino masses  discussed previously \cite{Beacom2004,Hannestad2005,Chacko2020,EscuderoRelaxing,Esteban2021,Barenboim2021,abellan2021, FrancoAbellan:2026ori} and offer some of the most promising avenues to reduce the Hubble tension \cite{Bell2006,Archidiacono2014,CyrRacine2014,Lancaster_2017,Park2019,Escudero:2019gvw,Kreisch2020,knox2020hubble,Choudhury2021,Venzor:2023aka}. The mass of the interaction mediator largely determines the effects of the 
interaction in the evolution of the neutrinos in the Universe. In particular, for light scalar mediators the interaction effects grow stronger at later times \cite{Forastieri2015,
Forastieri:2019cuf, Venzor:2022hql}, suppressing neutrino free streaming. This regime also induces 
an effective mass for neutrinos, which has been studied in high density environments, such as the early 
Universe \cite{Venzor:2020ova} or the cores of stars \cite{Ge2019,Babu2020scalar},as well as in high-energy astrophysical neutrino propagation \cite{Verma:2026wrs}.

Due to the above issues, the question of the correctness of the dark energy model and the 
neutrino mass constraints are very relevant research topics in cosmology today. For this reason we consider an alternative dark energy model that in turn is related 
to the origin of the neutrino mass.
To relate both phenomena, we will work in the framework of Unimodular Gravity \cite{Einstein:1919gv, ellis2011trace, Ellis:2013uxa}, a modification of the theory of General Relativity  where the vacuum energy has an associated
current density that appears from the non-conservation of the energy-momentum tensor. 
Different approaches within Unimodular Gravity have been proposed to explain the origin and dynamics of dark energy. These include scenarios in which energy-momentum non-conservation arises from microscopic quantum effects \cite{Josset:2016vrq,Perez:2017krv,Perez:2018wlo}, as well as effective cosmological models based on diffusion-like energy exchange between matter and dark energy sectors \cite{corral2020diffusion,LinaresCedeno:2020uxx,Landau:2022mhm}. In contrast, the mechanism considered here is driven by non-standard neutrino--scalar interactions, which transfer energy to an effective dynamical dark energy component.

This work is organized as follows. In Section \ref{foundations}, we describe our neutrino mass model, the 
origin of the dark energy model within Unimodular Gravity, and the physical assumptions adopted. In 
Section \ref{cosmo_constraints}, we present the datasets used for the Bayesian inference and the 
corresponding results for the one-neutrino and two-neutrino dark energy models. Finally, in Sections \ref{discussion} 
and \ref{conclusions}, we discuss our findings and summarize our conclusions.

\end{section} 

\section{Foundations of the model} \label{foundations}
\subsection{Neutrino mass model}
The basic idea underlying our model is that dark energy emerges from non standard interactions among the relic neutrinos in the late Universe, mediated by a light scalar field ($m_\phi<\mathcal{O}(\rm{eV})$). Such interactions induce finite-temperature corrections to the neutrino mass through forward-scattering processes in the cosmic medium. Physically, this mechanism can be interpreted as a modification of the neutrino dispersion relation in a thermal bath, analogous to medium-induced self-energy effects in finite-temperature field theory \cite{Weldon:1982bn,NotzoldRaffelt1988}. 

Within this scenario, neutrinos propagate in an environment where their effective mass receives contributions from both neutrino self-interactions and interactions with other neutrino species mediated by a scalar field $\phi$. These corrections depend explicitly on the temperature and the phase-space distribution of the background particles, leading to a dynamical mass that evolves with the expansion of the Universe. As a consequence, the neutrino sector becomes sensitive to the thermal history, and its backreaction can effectively source a time dependent dark energy component. 

In this work we will assume that the coupling is diagonal in the mass basis, with a strength $g_i$ dependent on the mass eigenstate $i$. To leading order in the couplings, the effective neutrino mass associated with the eigenstate $j$ is given by \cite{Babu2020scalar,Venzor:2020ova}:
\begin{equation}
m_{{\rm eff},j} =
m_j
+
\sum_{i=1}^3 G_{s,ij}\,\Delta\mathfrak{m}(m_i,T) \, ,\label{mef}
\end{equation}
where $m_j$ denotes the vacuum mass of the $j$-th neutrino species. The effective couplings
\begin{equation}
G_{s,ij}=\frac{g_i g_j}{m_\phi^2},
\label{effective_coup1}
\end{equation}
determine the strength of the interaction. We emphasize that, throughout this paper, commas appearing in subscripts serve only as label separators and do not denote partial differentiation.

The quantity $\Delta\mathfrak{m}(m_i,T_\nu)$ in Eq.~\eqref{mef} is the thermal
scalar density $\langle\bar{\nu}_i\nu_i\rangle$ of the bath species $i$,
including the contributions of neutrinos and antineutrinos:
\begin{equation}
    \Delta \mathfrak{m}(m_i,T) = \frac{m_i}{\pi^2}\int^{\infty}_0 dp \frac{p^2}{\sqrt{p^2+m_i^2}} \, f_i\,(p,T)\,, \label{deltam}
\end{equation}
where $p$ denotes the particle momentum and $f_i$ its distribution function.
Since neutrinos decouple in the early Universe, their distribution preserves the massless Fermi-Dirac form
\begin{equation}
f_i(p,T_\nu)=\frac{1}{e^{p/T_\nu}+1}\,,
\label{f_p}
\end{equation}
which is valid even when the neutrinos become non-relativistic.
The neutrino temperature evolves as $T_\nu(a)=T_{\nu,0}/a$,  with
$T_{\nu,0}=(4/11)^{1/3}\,T_{\gamma,0}\approx 1.68\times 10^{-4}\rm{eV}$.  $\Delta\mathfrak{m}$ inherits its time dependence directly from $T_\nu(a)$.

We are interested in linking the neutrino mass evolution to the dark energy of the Universe. The epoch of interest in the dynamics of the dark energy corresponds to the recent Universe, 
namely for $z \lesssim  2$. In the following sections we will see that species that are deep in the non-relativistic regime have a negligible effect on the dark energy dynamics. We therefore concentrate on the two lightest mass eigenstates: the lightest, which may be relativistic today and the second lightest which became non-relativistic at $z\sim\mathcal{O}(10)$. This also explains why we ignore contributions from the thermal baths of the charged leptons in \eqref{mef} which are much heavier, in contrast to Ref. \cite{Venzor:2020ova}.

\subsection{Origin of the \texorpdfstring{$\Lambda$}--term in the Einstein Equations}
In standard General Relativity, the cosmological constant is introduced as an independent parameter. In the present framework, however, we investigate the possibility that an effective dark energy component may arise from the energy transferred through the neutrino interactions described above. To implement this scenario, we adopt a framework in which the energy--momentum tensor is not strictly covariantly conserved.

Specifically, we employ the Unimodular Gravity (UG) theory \cite{Einstein:1919gv, Ellis:2013uxa}, which is not invariant under general diffeomorphisms, but volume-preserving  transformations. This in turn can be interpreted as the partial violation of energy-momentum conservation, provided they are integrable, $dJ=0$, for a current to be specified.  In our case, this current will be given by neutrino interactions.     

The field equations of Unimodular gravity are: 
 
      \begin{equation}
        R_{\mu\nu}-\frac{1}{4}R\,g_{\mu\nu}=8 \pi G \left(T_{\mu\nu}-\frac{1}{4}T\,g_{\mu\nu} \right), \label{TFE}
    \end{equation}
where the right-hand side term is the trace-free energy momentum contribution. This equation can be rearranged as:  
\begin{equation}
        R_{\mu\nu}-\frac{1}{2}R\,g_{\mu\nu}+ \Lambda g_{\mu\nu} =8 \pi G T_{\mu\nu}\, ,  \label{TFE4}
    \end{equation}  
that looks like the General Relativistic field equation, but now \cite{Perez:2017krv}:  
\begin{equation}
    \Lambda (x^{\mu}) = \Lambda_0 + 8 \pi G \int \nabla^{\nu}T_{\mu\nu}dx'^{\mu} =  \Lambda_0 + \int J_{\mu}dx'^{\mu} \, , \label{Lambda1}
\end{equation}
where $\Lambda_0$ is a constant of integration and the conserved  current is given by $J_{\mu}= 8 \pi G \nabla^{\nu}T_{\mu\nu}$. 
This current can evolve depending on physical processes that locally violate energy-momentum conservation \cite{Perez:2017krv,Josset:2016vrq}.  As a consequence, the term that plays the role of the cosmological constant in  Eq. (\ref{Lambda1}) becomes an effective quantity that can vary in space and time.

\subsection{Neutrino current}
To derive an expression for the current $J_\mu$ and subsequently evaluate the integral leading to $\Lambda$ in Eq.~(\ref{Lambda1}), we consider the energy--momentum tensor of a fermionic species $i$. In kinetic theory, it can be written as \cite{DodelsonSchmidt2025}
\begin{equation}
T_{\mu\nu}
= - 
\frac{g_d}{(2\pi)^3}
\int
p_\mu p_\nu\,f(p)\,
\frac{d^3p}{p^0},
\end{equation}
where $f(p)$ is given by Eq. (\ref{f_p}), and $g_d$ denotes the number of internal degrees of freedom of the species.  By invoking the cosmological principle and assuming a purely temporal force $F_\mu=\dot{p}_\mu$, only the time component of the modified conservation equation remains relevant, while the spatial components vanish due to homogeneity and isotropy,  the expression can be written as \cite{Rezzolla:2013dea}: 
\begin{eqnarray}
    \nabla^{\nu}\mathbf{T}_{\mu\nu} &=&-\frac{g_d}{(2 \pi)^3} \nabla^{\nu}\int p_{\mu}p_{\nu} f(p)\frac{d^3p}{p^0}. \label{EM1} \,
\end{eqnarray}
Starting from (\ref{EM1}) the spacetime derivative can be interchanged with the momentum integral, since the spacetime dependence is contained in the distribution function ($f(x,p)$), one obtains
\begin{equation}
\nabla^{\nu}\mathbf{T}_{\mu\nu}=
-\frac{g_d}{(2 \pi)^3}
\int p_{\mu}p_{\nu}
\nabla^\nu f
\frac{d^3p}{p^0}.
\end{equation}

Using the Boltzmann equation,
\begin{equation}
p^\nu\nabla_\nu f
+
F_\nu
\frac{\partial f}{\partial p_\nu}=
0,
\end{equation}
the spacetime derivative of the distribution function can be rewritten in terms of a momentum-space derivative, yielding
\begin{equation}
\nabla^\nu \mathbf{T}_{\mu\nu}=
\frac{g_d}{(2\pi)^3}
\int
p_\mu F_\alpha
\frac{\partial f}{\partial p_\alpha}
\frac{d^3p}{p^0}.
\end{equation}
The latter expression is integrated by parts in momentum space. Assuming that the boundary term vanishes, one finds
\begin{equation}
\int
p_\mu F_\alpha
\frac{\partial f}{\partial p_\alpha}
\frac{d^3p}{p^0} = -
\int f
\frac{\partial \left(
p_\mu F_\alpha
\right)}{\partial p_\alpha}
\frac{d^3p}{p^0}.
\end{equation}
Substituting this result into the previous equation and performing the angular integration, ($d^3p=4\pi p^2dp$), leads to
\begin{equation}
J_\mu=\nabla^{\nu}\mathbf{T}_{\mu\nu}=
-\frac{4\pi g_d}{(2 \pi)^3}
\int_0^{\infty}
m p^2 F_{\mu}f(p)\frac{dp}{p^0},
 \end{equation}
which will be used in the subsequent analysis.
 
A particle whose mass varies along its trajectory experiences an effective four-force associated with the rate of change of its mass. Following this interpretation, we model the force acting on the fluid associated with the neutrino eigenstate $i$ as
\begin{equation}
F_{i,\mu}=-\dot{m}_{\rm{eff},i}u_{i,\mu} \,,
\label{force}
\end{equation}
where $u_\mu$ is the fluid four-velocity. Since the neutrino fluid is
homogeneous and at rest in the cosmological (comoving) frame, its
four-velocity has only a time component. The force in Eq.~(\ref{force}) is then
purely timelike and represents the energy exchange induced by the evolution
of the effective neutrino mass.

Summing the forces over all the neutrino eigenstates, the 00-component of Eq. (\ref{EM1}) becomes:
\begin{equation}
    J_0=  -\frac{4\pi g_d}{(2 \pi)^3} \int_0^{\infty} \sum_{i=1}^3 \dot{m}_{{\rm eff},i} (T)  p^2f(p)dp \, . \label{EM2}
\end{equation}
Each of the fluid eigenstates receives contributions from the other thermal baths, exchanging the
order of summation,
\begin{eqnarray}
\sum_{i} \dot m_{{\rm eff},i}
&=& \sum_{i}\sum_{j} G_{s,ij}\,\dot{\Delta\mathfrak{m}}(m_j,T_\nu) \\ 
&=& \sum_{j} G_{s_j}\,\dot{\Delta\mathfrak{m}}(m_j,T_\nu), \label{suma_derivadas}
\end{eqnarray}
where the effective coupling of bath eigenstate $j$ is obtained by summing
over the sourced species $i$,
\begin{equation}
G_{s_j} \equiv \sum_{i} G_{s,ij}
= \frac{\rm{g_j} \sum_i \rm g_i}{m_\phi^2}
\,.
\label{effective_coup}
\end{equation}

The sum \eqref{suma_derivadas} can be factored out of the momentum integral in \eqref{EM2} despite the implicit dependence of $\Delta\mathfrak{m}$ as an integral of the momentum in Eq. \eqref{deltam}. This happens because the integral in $\Delta \mathfrak{m}$ is carried over the momenta of the background bath, while the integral in Eq. (\ref{EM2}) runs over the momenta of the fluid  eigenstate $i$ and yields its number density.

We can compute the integral over the momentum in Eq. (\ref{EM2}) analytically as: 
\begin{equation}
  \int_0^\infty p^2f(p)dp =\frac{3\zeta(3)}{2}T^3 \, .\label{pUR}
\end{equation}
Evaluating $J_{0}=8 \pi G \nabla^{0}\mathbf{T}_{00}$ 
with Eqs. (\ref{mef}), (\ref{deltam}), (\ref{EM2}), and (\ref{pUR}) we obtain:
 \begin{eqnarray}
    J_0 &=&- (8 \pi G )
    \sum_{j=1}^3
    \frac{3 \zeta(3) g_d G_{s_j} m_j}{4 \pi^4} \nonumber \\
&&\times \int ^{\infty}_{0}
    \frac{T_{\nu} e^{p/T_{\nu}}p^3 \dot{T_{\nu}} dp}{\sqrt{p^2+m_j^2}(e^{p/T_{\nu}}+1)^2} , \label{J_R}
 \end{eqnarray}
where we used the fact that $\dot{m}_{{\rm eff},j}= (dm_{{\rm eff},j}/dT_\nu)\dot T_\nu$. The integral runs over the momenta and we will evaluate it numerically to obtain the evolution of the dark energy in the next section.

\subsection{Dark Energy from neutrino interactions}

To fully specify the dark energy model, the momentum integral appearing in the neutrino distribution function must be supplemented by the time integration entering Eq.~(\ref{Lambda1}). For the $00$ component, Eq.~(\ref{Lambda1}) becomes:
\begin{equation}
\Lambda(t) 
= \Lambda_0 + \int_{t_0}^{t} J_{0}\,dt' .
\label{Lambda_00}
\end{equation}

Using Eq.~(\ref{J_R}) and changing the integration variable from cosmic time to the neutrino temperature, $dt=dT_\nu/\dot T_\nu$, and the definition $\rho_{\Lambda} = \frac{\Lambda}{8 \pi G}$ :
\begin{equation}
\begin{split}
\rho_{\Lambda}(T_{\nu})
&= \rho_{\Lambda}(T_{\nu,0})
-\frac{3\zeta(3)\, g_d \, }{4\pi^4}
\sum_{j=1}^{3}
G_{s_j} \, m_j
\int_{T_{0,\nu}}^{T_{\nu}}
\Bigg[
dT'_{\nu} \, T'_{\nu} \\
&\quad \times \int_{0}^{\infty}
\frac{e^{p/T'_{\nu}}\, p^3\, dp}
{\sqrt{p^2+m_j^2}\left(e^{p/T'_{\nu}}+1\right)^2}
\Bigg]
\, ,
\end{split}
\label{rho_T}
\end{equation} 
where
\begin{equation}
\rho_{\Lambda,0}\equiv \rho_{\Lambda}(T_{\nu,0})
=\frac{\Lambda_0}{8\pi G}, \label{def_rho_0}
\end{equation}
is an integration constant corresponding to the dark energy density evaluated at the reference neutrino temperature ($T_{\nu,0}$). Throughout this work, we identify ($T_{\nu,0}$) with the present-day neutrino temperature, ($T_\nu(z=0)$). The sum runs over the three neutrino mass eigenstates, (i=1,2,3). The condition ($T_{f,\nu}>T_{0,\nu}$) indicates that the integration extends from the present epoch ($z=0$) to earlier times characterized by higher neutrino temperatures and larger redshifts. Since the integral contribution in Eq.~(\ref{rho_T}) grows monotonically with ($T_{f,\nu}$), sufficiently large values of the upper integration limit would eventually drive ($\rho_\Lambda$) to negative values. To ensure a positive dark energy density, we assume that the neutrino--scalar interaction is effective only up to a characteristic temperature ($T_{\nu,\, H}$). Above this temperature the interaction is assumed to switch off, and therefore no further contribution to ($\rho_\Lambda$) is generated. 

To gain insight into the evolution of $\rho_\Lambda$, 
we compute the integrals in \eqref{rho_T} at the ultrarelativistic and non-relativistic level.
The definite integral appearing in Eq.~($\ref{rho_T}$) is evaluated with respect to a reference temperature. As a result, the dependence on the integration limits ($T_{\nu,0}$) and ($T_{\nu}$) is absorbed into the corresponding integration constants. Consequently, the quantities ($\rho_\Lambda(T_t)$) and ($\rho_\Lambda(T_{\nu,0})$) appearing in the ultra-relativistic and non-relativistic solutions, respectively, represent the value of the dark energy density at the chosen reference temperature.

For neutrinos in their ultra-relativistic regime ($T=T_{\nu}\gg m_j$) $\rho_\Lambda$ evolves as:
\begin{align}
\rho_{\Lambda_{UR}}(T)
&=
\rho_{\Lambda}(T_t)
-\frac{3\,\zeta(3)\,g_d}{4\pi^4}
\sum_j G_{s_j}m_j
\nonumber\\
&\times
\Bigg[
\frac{\ln(2)\,m_j}{4}
\left(T^{4}-T_t^{4}\right)
+\frac{\pi^{2}}{30}
\left(T^{5}-T_t^{5}\right)
\Bigg].
\label{rho_ur}
\end{align}
where $T_t$ is some reference temperature.

On the other hand, for neutrinos in their non-relativistic regime ($T=T_{\nu} \ll m_j$) Eq. (\ref{rho_T}) becomes: 
\begin{align}
\rho_{\Lambda_{NR}}(T)
&=
\rho_{\Lambda}(T)
+\frac{g_d}{(2\pi)^4}
\sum_j G_{s_j} m_j^3
\nonumber\\
&\times
\Big[
T^{3}e^{-2m_j/T}-T_{\nu,0}^{3}e^{-2m_j/T_{\nu,0}}
\Big].
\label{rho_nr}
\end{align}
where
$T_{\nu,0}$ is the present-day neutrino temperature.
The quantity $\rho_\Lambda(T_{\nu,0})$ corresponds to the
integration constant fixed by the value of Eq.~(\ref{rho_T})
at the present epoch. Therefore, in the limit
$T_{\nu,L}\to 0$, the temperature-dependent contribution is
exponentially suppressed and the dark energy density tends to
the constant value $\rho_\Lambda(T_{\nu,0})$.

The time (temperature) dependence  of $\rho_\Lambda$ decays rapidly for fluids in the non-relativistic regime. For different neutrino species, the ones in the non-relativistic limit therefore stop contributing to the evolution of the dark energy. If we are interested in modeling an evolving dark energy, we need to focus only on neutrinos in the relativistic limit or that have recently transitioned to the non-relativistic regime.
In the physical scenarios considered below, the dark energy evolution is driven mainly by the relativistic to non-relativistic transition of the lightest neutrino species, occurring around $T_\nu \sim m_1$. At this stage, the neutrino distribution departs from its ultra relativistic behavior and the interaction-induced contribution to $\rho_\Lambda$ evolves most efficiently, giving rise to the dynamical dark energy behavior explored in this work.

\subsection{Neutrino mass and models of dark energy density} \label{neutrino_beha}

We analyze two cases: first, considering a coupling only to the lightest
eigenstate, $m_1$; and second, considering couplings to the two lightest
eigenstates, $m_1$ and $m_2$. In the first case the neutrino mass ordering
plays no role in the interacting sector; in the second case, however, we
want the second-lightest eigenstate to be as light as possible, so that it
remains close to the relativistic regime until recent times. We therefore
adopt the normal mass ordering in this work. In this ordering the mass of the second-lightest neutrino is bounded from below by the solar neutrino experiments,
$m_2 \geq \sqrt{\Delta m_{21}^2} \approx 8.7\times10^{-3}\,\mathrm{eV}$,
whereas in the inverted ordering it would lie at the
atmospheric neutrino constraint, $\sim 5\times10^{-2}\,\mathrm{eV}$, and would have become
non-relativistic earlier. In this case, once the mass of the lightest
neutrino eigenstate, $m_1$, is specified, the masses of the remaining
eigenstates are determined through
 \begin{equation}
    m_2 = \sqrt{\Delta m_{21}^2 + m_1^2} \, , \label{Deltam2}
\end{equation}
\begin{equation}
    m_3 = \sqrt{\Delta m_{31}^2 + m_1^2} \, , \label{Deltam3}
\end{equation}
where $\Delta m_{21}^2 = 7.49 \times 10^{-5}\mathrm{eV}^2$ and
$\Delta m_{31}^2 = 2.48 \times 10^{-3}\mathrm{eV}^2$
\cite{Gariazzo:2018pei, Long:2017dru, RoyChoudhury:2018gay,
Hannestad:2016fog, Loureiro:2019hwh}.

We will determine the range of the lightest neutrino mass $m_1$ motivated by the epoch at which the neutrino transitions from the ultra-relativistic to the non-relativistic regime, which is the regime of primary interest in our model. This transition occurs when $m_1 = \langle p \rangle \approx 3.15\,T_\nu(z)$, so that only masses above $3.15\,T_{\nu,0} \approx 5.3\times10^{-4}\,\mathrm{eV}$ have
become non-relativistic by the present epoch. We therefore consider
representative values in the interval
$5\times10^{-5}\,{\rm eV}\le m_1\le10^{-3}\,{\rm eV}$: the lightest cases
remain relativistic today, while the heaviest become non-relativistic in
the recent past, at $z \lesssim 0.9$.

Cosmological observations constrain the total neutrino mass and therefore indirectly restrict the lightest mass eigenstate. The interval adopted
here lies well below current cosmological bounds on the lightest eigenstate $m_1 < 0.02\,\mathrm{eV}$ (95\% C.L.) obtained DESI dataset \cite{Chebat:2025kes, Elbers:2025vlz}. We note that this
bound is derived under standard cosmological assumptions and relaxes
considerably in dynamical dark energy models such as the one considered
here.

Since the dynamical dark energy component in our model is driven by the evolution of the neutrino thermal correction across this transition, this interval allows us to explore the regime where observable departures from $\Lambda$CDM may arise. Interestingly, recent DESI analyses \cite{DESI:2024mwx,DESI:2025zgx,DESI:2025fii} indicate a preference for an evolving dark energy at low redshift, with a maximum density around $z\sim0.6$, which lies within the redshift range covered by the neutrino masses considered here. Conversely, when the neutrino-induced
thermal corrections become negligible, Eq.~(\ref{mef}) reduces to the
standard neutrino mass spectrum and the cosmological evolution recovers
the $\Lambda$CDM limit.

\subsubsection{Single-neutrino model}
For the numerical analysis, we employ the full expression for the dark energy density given by  Eq.~(\ref{rho_T}) and restrict the discussion to the lightest neutrino mass eigenstate, making the coupling $g_1\neq 0$ while $g_2=g_3=0$.
This model corresponds to a minimal scenario in which only the lightest 
neutrino species contributes to the effective dark energy sector. As a consequence, the resulting energy 
density evolution is controlled by a single mass scale and a single coupling parameter. It is our simplest scenario that limits the  
dynamical richness of the model. 

From a phenomenological perspective, the evolution of $\rho_\Lambda(T)$ is smooth and monotonic. This behavior follows directly from the dynamical structure of the model. In particular, Eq.~(\ref{Lambda_00}) implies that $\dot{\Lambda}=J^0$, while Eq.~(\ref{J_R}) fixes the sign of $J^0$ throughout the cosmological evolution, thereby enforcing a monotonic evolution of $\Lambda$. As the neutrino temperature decreases with the expansion of the Universe, the phase-space suppression encoded in the Fermi-Dirac distribution progressively reduces the interaction contribution. Consequently, the neutrino induced dark energy density increases toward its present-day value and gradually freezes at late times, asymptotically approaching a constant. This behavior is a generic prediction of the model and is determined by the form of the neutrino dark energy coupling. In particular, the opposite evolution, corresponding to a dark energy density decreasing from larger values toward $\rho_\Lambda$, cannot be realized within the physically allowed parameter space, as it would
require the effective coupling to be negative. However
$G_{s_1}=g_1^2/m_\phi^2$ is manifestly positive. A sign change becomes possible only in the presence of multiple couplings, where Eq.~\eqref{effective_coup} is no longer a perfect square.

\subsubsection{Two-neutrino model}
In this case,  we consider the two lightest neutrinos, setting different values for $m_1$  and the other given by Eq. \eqref{Deltam2}. Each neutrino has its $G_s$-coupling, then using Eq. (\ref{rho_T}) for two-neutrino,  we have:
\begin{widetext}
\begin{eqnarray}
    \rho_{\Lambda}(T_{\nu})= \rho_{\Lambda}(T_{\nu,0})  -    \frac{ 3 \zeta(3)g_{\nu}}{ 4 \pi ^4} \,\left[ G_{s_1} \, m_1   \int_{T_{0,\nu}}^{T_{\nu}}\left( T'_{\nu}\int ^{\infty}_{0}
    \frac{e^{p/T'_{\nu}}p^3dp}{\sqrt{p^2+m_1^2}(e^{p/T'_{\nu}}+1)^2}\right) dT'_{\nu}\right.  \nonumber \\
+\left. G_{s_2} \, m_{2}  \int_{T_{0,\nu}}^{T_{\nu}}\left(T'_{\nu}\int ^{\infty}_{0}
    \frac{e^{p/T'_{\nu}}p^3dp}{\sqrt{p^2+m_{2}^2}(e^{p/T'_{\nu}}+1)^2}\right)  dT'_{\nu} \right], \,  \label{rho_T12}
\end{eqnarray}
\end{widetext}
To improve the convergence of the sample method in subsection \ref{inference2}, we define 
\begin{eqnarray}
    \eta \equiv \frac{G_{s_1}}{G_{s_2}}, \label{eta}
\end{eqnarray}
and vary $(G_{s_2},\eta)$ as independent parameters, with $G_{s_1}$ derived.

This two-neutrino scenario introduces a non-trivial interplay between the contributions of each species. We will see that the dataset prefers opposite signs for the two couplings. As a consequence, the dark energy density $\rho_\Lambda(T)$ is no longer monotonic, but instead exhibits a richer dynamical structure arising from the partial cancellation between the two terms.

In particular, the definition of  $ \eta $  induces a competition between the two neutrino contributions, each weighted by a different mass scale. At early times (high $T_\nu$), both species contribute, and the combined effect can lead to an enhancement of $\rho_\Lambda$. However, as the Universe expands and $T_\nu$ decreases, the heavier neutrino becomes Boltzmann suppressed earlier than the lighter one, breaking the balance between the two terms.

If $G_{s_1}<0$ and $G_{s_2}>0$, this imbalance naturally gives rise to a peaked behavior in the evolution of the dark energy density, where $\rho_\Lambda$ grows up to a maximum at intermediate redshifts and subsequently decreases toward late times. Such a feature resembles a transient dark energy component and departs noticeably 
from the standard $\Lambda$CDM expectation of a constant energy density.

From a phenomenological point of view, this behavior may alleviate tensions in observational dataset by allowing for a dynamical dark energy sector that modifies the expansion rate in a redshift-dependent manner. In particular, the presence of a maximum in $\rho_\Lambda(z)$ around $z \sim \mathcal{O}(1)$  \cite{DESI:2025fii}, can impact distance measurements and the growth of structure, providing a potential signature to constrain the model.

Finally, the parameter $\eta$ controls the relative strength of the cancellation and therefore regulates both the amplitude and the position of the peak in $\rho_\Lambda$. This introduces an additional degree of freedom that enhances the flexibility of the model, at the cost of increasing parameter degeneracies when fitting to cosmological data. Recent DESI observations \cite{DESI:2025zgx}, when combined with other cosmological probes, have renewed interest in dark energy scenarios that evolve with redshift rather than remaining strictly constant. In this context, we consider a phenomenological framework in which the interaction between neutrinos and dark energy allows the effective dark energy density to increase from the past, attain a maximum value, and then decrease at late times. This possibility motivates the sign conventions adopted for the interaction parameters and provides a concrete observationally inspired scenario to test against the data.

\section{Cosmological constraints}\label{cosmo_constraints}

To assess both models, we confront them with different observational datasets by exploring a range of values 
for the lightest neutrino mass. This procedure allows us to constrain the coupling $G_{s_1}$ within the single-neutrino framework described by Eq.~(\ref{rho_T}). We then extend the analysis to the two-neutrino scenario, Eq.~(\ref{rho_T12}), where the interaction of the intermediate-mass neutrino $m_2$ is characterized by the coupling $G_{s_2}$. The coupling associated with the lightest neutrino, $G_{s_1}$, is subsequently determined through $\eta$, Eq. (\ref{eta}).

\subsection{Datasets}
To constrain the cosmological parameters of the models, we combine several complementary late- and early-Universe probes spanning the redshift range from the local Universe up to recombination. Specifically, we consider Megamasers, Cosmic Chronometers, SNe Ia DES-Dovekie, BAO DESI DR2, and the CMB angular acoustic scale $\theta_*$ measured by Planck. These datasets  provide independent constraints on the cosmic expansion history, cosmological distances, and the 
sound horizon scale, allowing us to test the consistency of the models across different epochs and 
observational techniques. In all analyses, we include the corresponding covariance matrices and systematic 
uncertainties provided by each collaboration.

\begin{enumerate}

\item \textbf{Megamasers.} Geometric distance measurements from water Megamaser (MM) galaxies observed by the Megamaser Cosmology Project \cite{Pesce_2020_Megamasers} are incorporated in the analysis. The sample contains six Megamaser galaxies, providing angular diameter distances and recession velocities in the low redshift range $0.002\leq z\leq 0.034$. They provide a direct late Universe constraint on $H_0$.

\item \textbf{Cosmic Chronometers.} We employ cosmic chronometer (CC) measurements of the expansion rate $H(z)$, obtained from the differential age evolution of passively evolving massive galaxies. This method provides direct estimates of the Hubble parameter through $H(z)=-(1+z)^{-1}dz/dt$, with minimal cosmological assumptions since it relies on relative ages rather than absolute age calibration. The compilation contains 33 measurements in the range $0.07\leq z< 2$, collected from several surveys and analyses \cite{Moresco:2020fbm}, including the full covariance matrix following the prescription of \cite{Moresco_Covariance}. These dataset help break degeneracies between $H_0$ and the absolute magnitude $M_{\rm B}$ entering the Supernovae analysis, as well as between $H_0$ and the sound horizon scale $r_d$ appearing in BAO measurements.
\item \textbf{Type Ia supernovae.} We use the most recent Type Ia Supernova (SNe Ia) compilation from the DES Dovekie reanalysis \cite{10.1093/mnras/stag632_DES_Dovekie} of the DES 5 year supernova sample \cite{Abbott_2024_DES_old},  which provides  standardized apparent magnitudes, $m_B$, for the SNe Ia together with the corresponding covariance matrix. This reanalysis incorporates an updated photometric cross calibration, new white dwarf calibration data, a retrained SALT3.DOV light curve model, regenerated simulations, corrected calibration uncertainties, and an exact implementation of the Fitzpatrick color law. The final DES Dovekie cosmology sample includes 1623 likely DES SNe Ia together with 197 low redshift SNe Ia, over the redshift range $0.025 \lesssim z \lesssim 1.1$. 

\item \textbf{DESI DR2 BAO.} We use baryon acoustic oscillation (BAO) measurements from DESI DR2 \cite{DESI:2025zgx}, based on galaxy, quasar, and Ly$\alpha$ forest tracers. These dataset constrain cosmological distances relative to the sound horizon $r_d$, namely $D_M/r_d$, $D_H/r_d$, and the angle averaged distance $D_V/r_d$. The effective redshifts are $z_{\rm eff}=0.295,0.510,0.706,0.934,1.321,1.484,2.330$, covering the range from low redshift galaxies to Ly$\alpha$ forest absorption.

\item \textbf{Planck distance priors.} Instead of using the full CMB likelihood, we adopt the Planck 2018 distance priors from the final TT,TE,EE+lowE release \cite{Chen_2019_Planck_DP}. These compressed CMB constraints encode the main geometrical information relevant for late Universe expansion through the acoustic scale $l_A$, the shift parameter $R$, and $\Omega_bh^2$, including their covariance. The quantities $l_A$ and $R$ characterize the angular scale of the acoustic peaks and the distance to the photon decoupling epoch, respectively, and are computed from the comoving sound horizon and angular diameter distance at recombination. This approach provides an efficient approximation to the full CMB likelihood while retaining most of the constraining power relevant for dark energy models. In the flat $\Lambda$CDM case, the values are $R=1.7502\pm0.0046$, $l_A=301.471^{+0.089}_{-0.090}$, and $\Omega_bh^2=0.02236\pm0.00015$. Their uncertainties are about $40\%$ smaller than those from Planck 2015 TT+lowP, and they reproduce the full Planck 2018 constraints for several dark energy models with good accuracy. We will refer to this dataset as $\theta_\ast$.  
\end{enumerate}

In the following sections, we will refer to the combination MM + CC + SNe Ia (DES-Dovekie) + DESI BAO DR2 as the \emph{low-redshift probes} or simply (\textbf{LzP}) data. When including the Planck distance priors, \emph{high-redshift probe}, we denote the dataset combination as (\textbf{LzP}+${\theta_*})$ data. 

All likelihood analyses presented in this work were performed using a modified version of the Late Universe dataset Binning (LUDB) code \footnote{See \url{https://github.com/MauLoHdz/LUDB}}, originally developed for cosmological parameter inference from late Universe observations. The code implements the likelihood functions for the observational datasets described above, including the corresponding covariance matrices and systematic uncertainties. It provides a flexible framework for combining different datasets, evaluating the total $\chi^2$ for a given cosmological model, and performing cosmological parameter inference through Markov Chain Monte Carlo (MCMC) sampling. For the present analysis, the original implementation was extended to incorporate the interacting models considered in this work through modifications of the cosmological background evolution and the corresponding theoretical predictions entering the likelihood calculations.

\subsection{Cosmological inference for the one-neutrino model}
\begin{figure}[t!]
\centering
\includegraphics[scale=0.5]{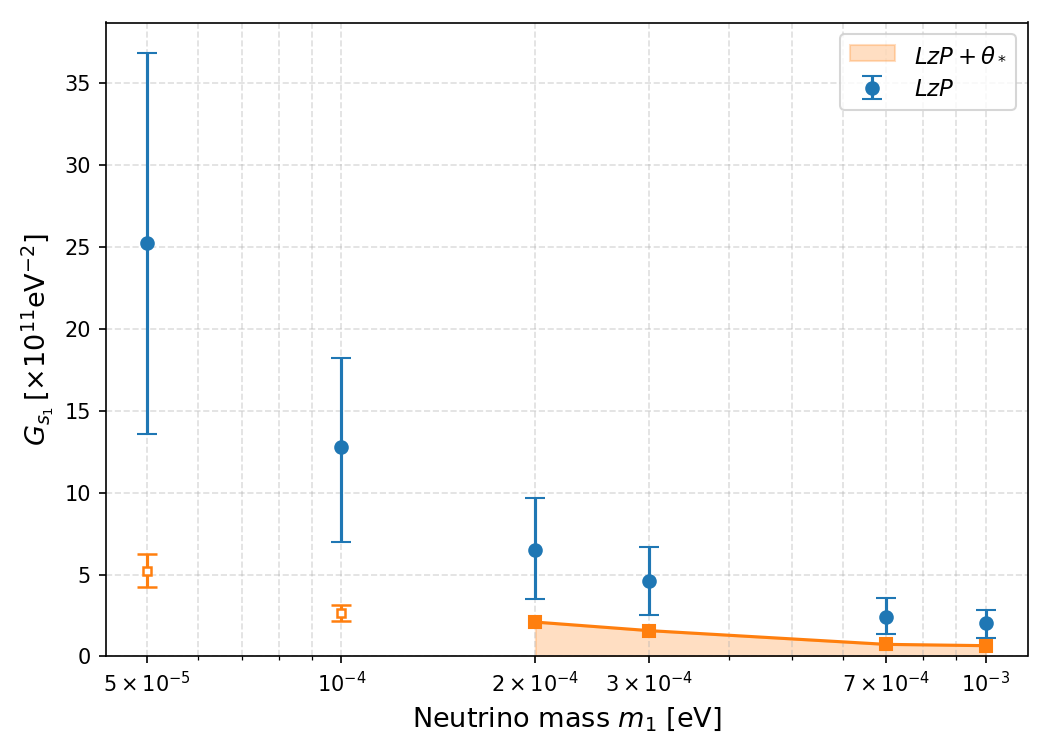}
\caption{Values for $G_{s_1}$ (in blue) obtained from the following datasets:  Megamasers, Cosmic Chronometers, SNe Ia DES-Dovekie, and DESI BAO DR2, i.e., \textbf{LzP}. Whereas $G_{s_1}$ in orange uses the same data as before, but adding Planck CMB distance priors \textbf{LzP}+ $\theta_*$. The shaded orange region marks the 1$\sigma$ limit allowed by the dataset.}
\label{DD_all_chart}
\end{figure}

As we mentioned above, we consider a range of  $m_\nu$ values to determine $G_{s_1}$, from $5.0\times10^{-5}$eV to $1.0\times10^{-3}$eV, since in this interval the cosmological dynamics has an interesting effect for dark energy, cf. Section \ref{neutrino_beha}.  We employ a flat prior for the interaction strength,  $G_{s_1} \in [0, 100] \times 10^{11} \rm eV^{-2}$, a range that proves convenient for the cosmological dynamics.
In Table~\ref{G-val_1n_t} we report the values of $G_s$ that we obtained,
where we identify two main features. As the neutrino mass increases, $G_{s_1}$ with \textbf{LzP} dataset decreases in inverse proportion to $m_1$, in accordance with the coupling $G_{s_1}  m_1$ in Eq. (\ref{rho_T}). Moreover, the inclusion of CMB priors  (\textbf{LzP}  + $\theta_*$) in the analysis further diminishes $G_{s_1}$; it tends to approach zero for larger neutrino masses. This trend is illustrated graphically in Fig. \ref{DD_all_chart}. 

\begin{table*}[t]
\centering
\small
\setlength{\tabcolsep}{7pt}
\begin{tabular}{c c c c c c c}
\hline
$m_1$ [eV]
& $G_{s_1}$ (\textbf{LzP})
& $N_\sigma$
& $G_{s_1}$ (\textbf{LzP}+$\theta_*$)
& $N_\sigma$
 \\
\hline
$5.0\times10^{-5}$
& $25.21^{+11.64}_{-11.64}$
& $1.95$

& $5.24^{+4.19}_{-2.88}$
& $1.47$
 \\
$1.0\times10^{-4}$
& $12.77^{+5.42}_{-5.80}$
& $2.01$

& $2.65^{+1.88}_{-1.43}$
& $1.60$
 \\
$2.0\times10^{-4}$
& $6.52^{+3.15}_{-3.00}$
& $2.00$

& $<2.1$
& ---
\\
$3.0\times10^{-4}$
& $4.62^{+2.08}_{-2.07}$
& $2.01$

& $<1.58$
& ---
 \\
$7.0\times10^{-4}$
& $2.44^{+1.13}_{-1.08}$
& $1.97$

& $<0.74$
& ---
 \\
$1.0\times10^{-3}$
& $2.03^{+0.82}_{-0.92}$
& $2.1$

& $<0.66$
& ---
 \\
\hline
\end{tabular}
\caption{Constraints on the effective coupling $G_{s_1}\,[\times 10^{11}\,\mathrm{eV}^{-2}]$ for different values of the lightest neutrino mass $m_1$ from the LzP and LzP+$\theta_*$ analyses. Uncertainties correspond to the 68\% credible interval, while upper limits indicate no statistically significant detection. The quantity $N_\sigma$ denotes the  significance of the preference for the best-fit model over the null hypothesis $G_{s_1}=0$.} 
\label{G-val_1n_t}
\end{table*}

As shown in Table~\ref{G-val_1n_t}, the \textbf{LzP} dataset alone shows a preference for positive, nonzero values of $G_{s_1}$. The quantity $N_\sigma$ denotes the Gaussian-equivalent significance of the preference for the best-fit model over the null hypothesis $G_{s_1}=0$, obtained from the profile likelihood via $\Delta \chi^2 = \chi^2(G_{s_1}=0) - \chi^2_{\min}$. This quantity provides a rough estimate of the statistical preference against the null-coupling hypothesis, $G_{s_1}=0$. Accordingly, the marginalized central value lies approximately $2 \sigma$ away from the null-coupling case, $G_{s_1}=0$. This suggests that the null-coupling scenario provides a slightly poorer description of the data. However, the statistical significance remains modest and should not be interpreted as evidence for a nonvanishing coupling, but rather as a weak indication of a possible neutrino--dark energy interaction. The inclusion of the $\theta_\ast$ prior substantially weakens this preference, leading to smaller best-fit values and, for larger neutrino masses, only upper limits on $G_{s_1}$.

As we can see in Table \ref{G-val_1n_t}, among the neutrino masses explored, the case $m_1=5.0\times10^{-5}\mathrm{eV}$ produces the largest departure from the $\Lambda$CDM prediction. For this  case, we obtain the $2\sigma$ upper bound $G_{s_1}<4.83\times10^{12}\mathrm{eV}^{-2}$ from the {\bf LzP} data. Including the CMB acoustic-scale constraint $\theta_\ast$ further strengthens the limit to $G_{s_1}<1.77\times10^{12}\mathrm{eV}^{-2}$ at $2\sigma$.

\begin{figure}[t!]
\centering
\includegraphics[scale=0.5]{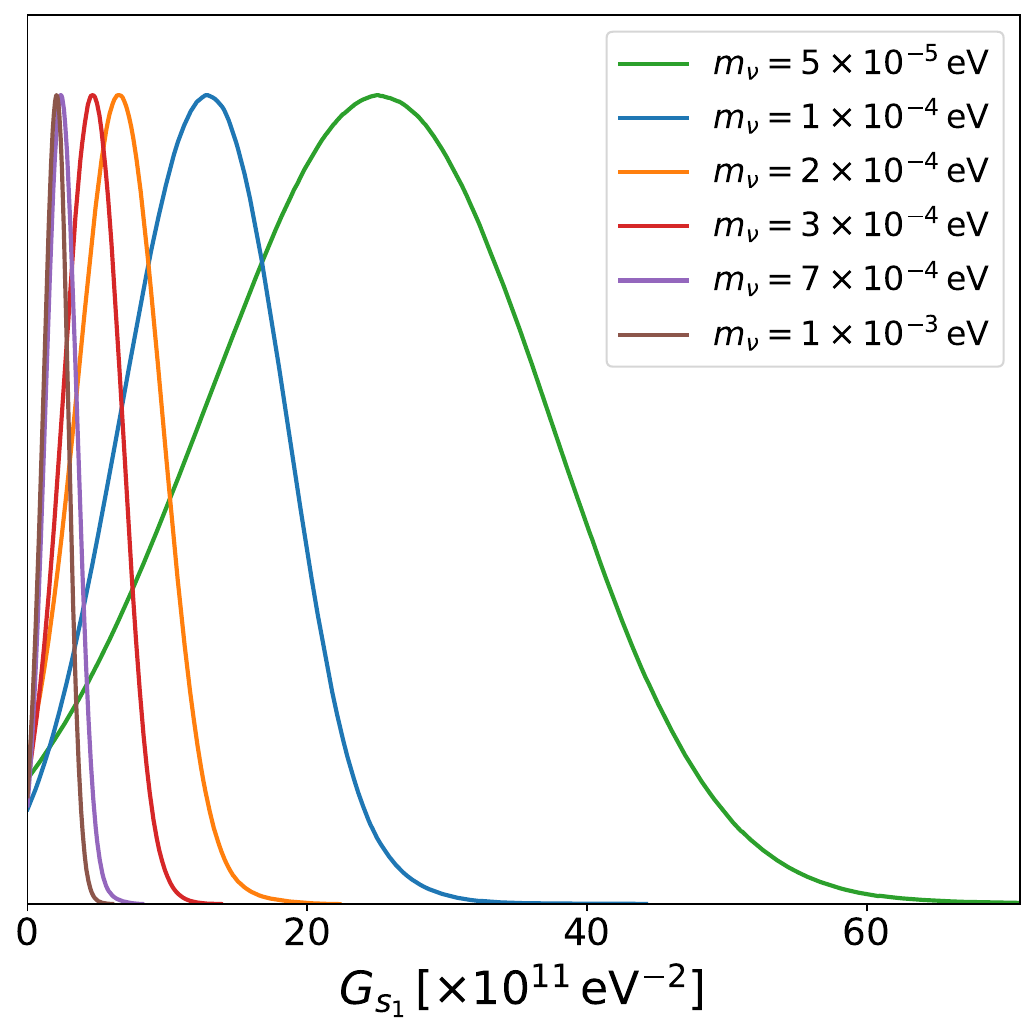}
\caption{Posterior distributions for $G_{s_1}$ values with different neutrino masses for the  \textbf{LzP} dataset. }
\label{DD_all_dist}
\end{figure}

Fig. \ref{DD_all_dist} shows the posterior distributions of $G_{s_1}$, confirming that
smaller neutrino masses allow larger deviations from $\Lambda$CDM.
Conversely, 
larger neutrino masses drive the interaction strength toward zero and therefore toward the $\Lambda$CDM limit. This behavior is strengthened by the inclusion of the $\theta_*$ prior, for which $G_{s_1}$ becomes consistent with zero at the $1\sigma$ level for masses greater than 0.2 meV. 

\begin{figure}[t!]
\centering
\includegraphics[scale=0.5]{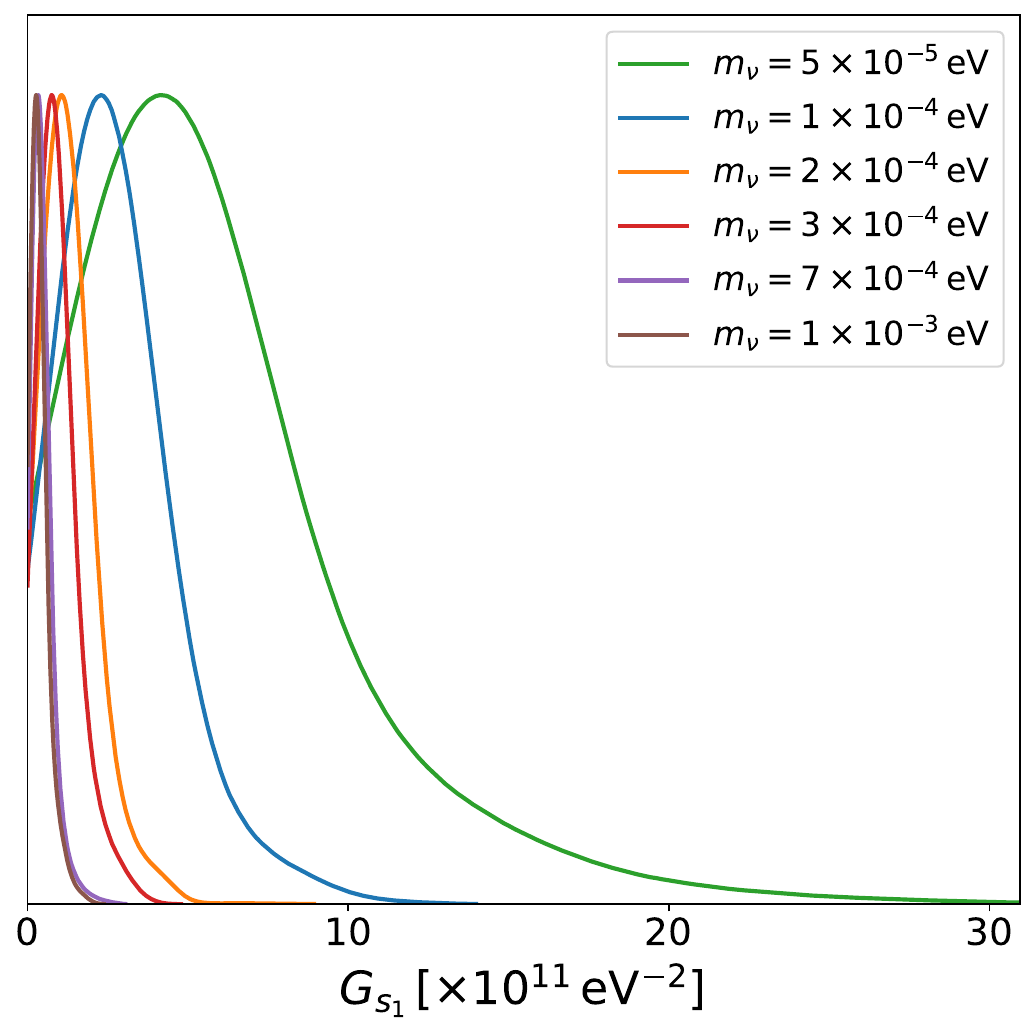}
\caption{Posterior distributions for $G_{s_1}$ with different neutrino masses for  the \textbf{LzP}+$\theta_*$ dataset. }
\label{DD_allPl_dist}
\end{figure}

As shown in Fig.~\ref{DD_all_dist} and \ref{DD_allPl_dist}, the largest deviation from the $\Lambda$CDM limit occurs for $m_\nu = 5\times10^{-5}\mathrm{eV}$ in both datasets. The effect of including the $\theta_*$ constraint from Planck is clearly illustrated when comparing the distributions, where a  narrowing and shift of the posterior distributions are observed. This reflects a reduction of the allowed parameter space once early Universe information is taken into account.

\begin{figure}[t!]
\centering
\includegraphics[scale=0.35]{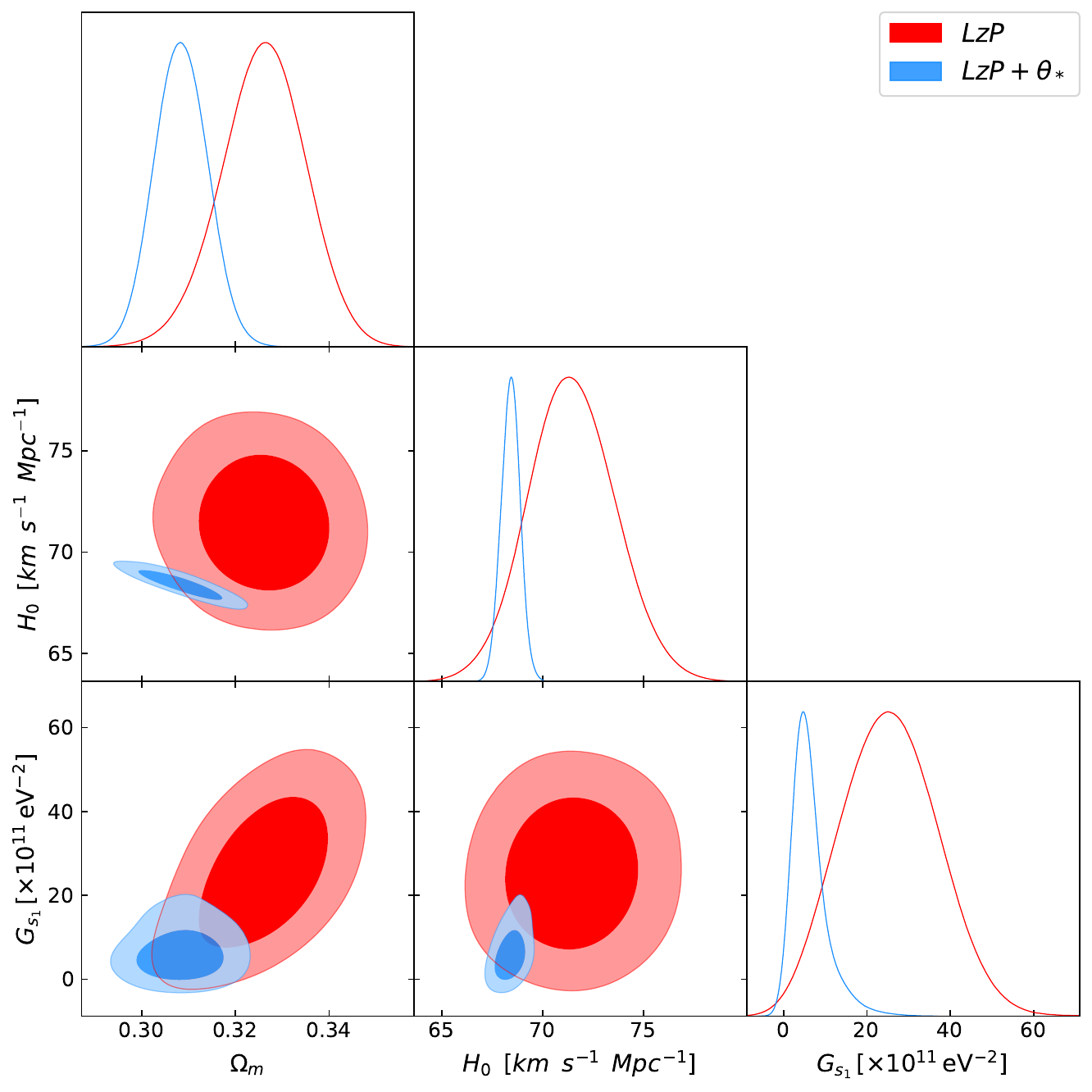}
\caption{\normalsize Comparison of the marginalized posterior distributions for $m_\nu = 5 \times 10^{-5}\,\mathrm{eV}$, which corresponds to the case that exhibits the largest deviation from $\Lambda$CDM among those studied. The analysis uses \textbf{LzP} dataset (red), and the same dataset combination including $\theta_*$ (blue). The shaded contours correspond to the $1\sigma$ and $2\sigma$ confidence levels. }
\label{5e-5_all}
\end{figure}

\begin{figure}[htbp]
\centering
\includegraphics[scale=0.35]{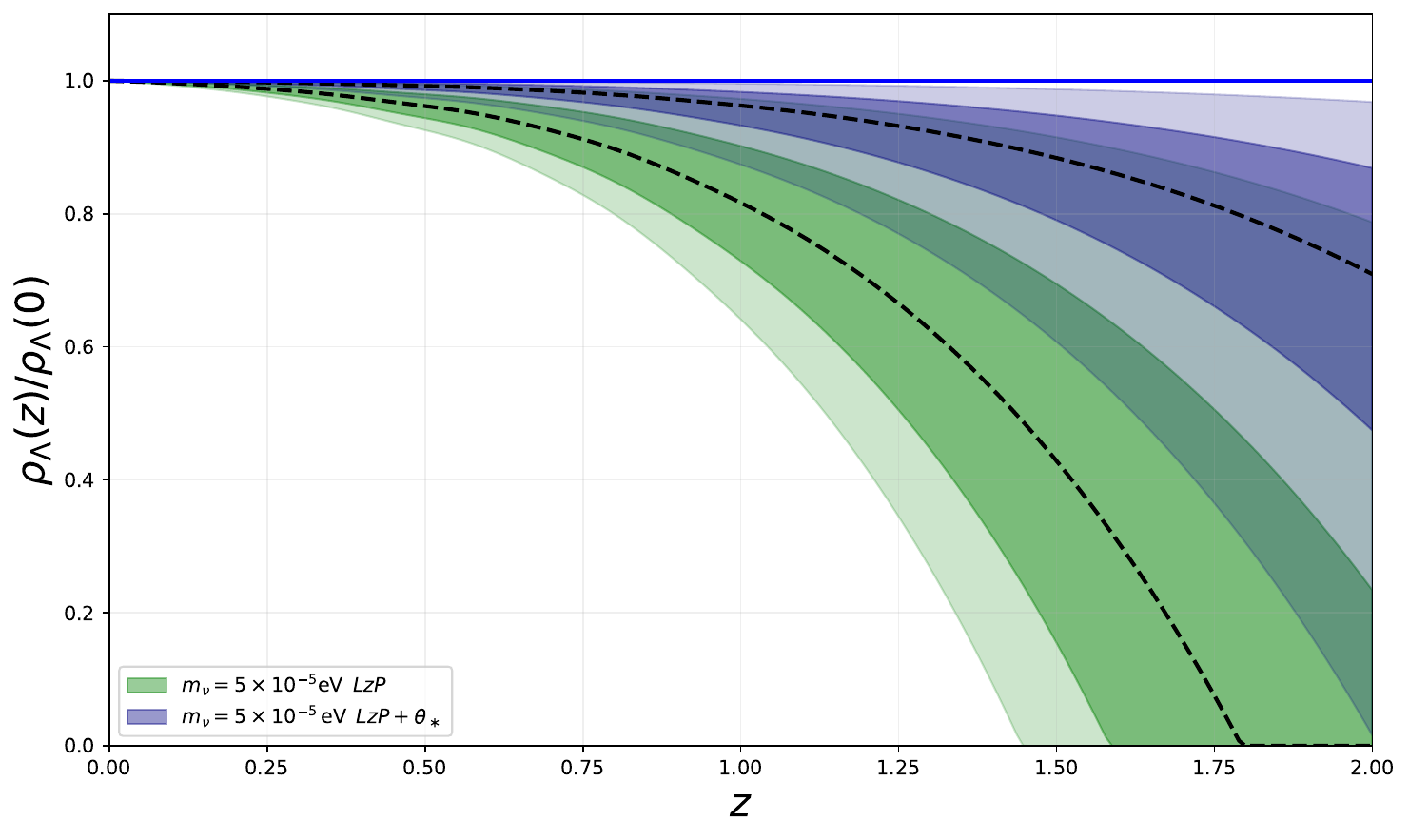}
\caption{Dark energy density evolution for $m_\nu=5\times 10^{-5} \rm{eV}$.Dark and light shaded bands indicate the $68\%$ and $95\%$ confidence intervals given the data, while dashed lines show the median reconstruction. The horizontal blue line denotes the $\Lambda$CDM prediction.}
\label{5e-5_3DD_rho_}
\end{figure}

Consistently, this impact is also visible in the two-dimensional contours  shown in Fig.~\ref{5e-5_all}, where the inclusion of $\theta_*$ leads to a shrinkage of the confidence regions.For illustrative purposes, we display this comparison only for the case $m_\nu = 5\times10^{-5},\mathrm{eV}$. Finally, Fig.~\ref{5e-5_3DD_rho_} shows the evolution of the normalized dark energy density as a function of redshift, $\rho_{\Lambda}(z)/\rho_{\Lambda}(0)$, obtained from the posterior probability, confirming that the tighter constraints translate into a more controlled and less dispersive cosmological evolution.

The evolution of the normalized dark energy density  provides key insights into the dynamical properties of the dark energy sector within the considered model. As shown in Fig. \ref{5e-5_3DD_rho_}, both reconstructions are normalized to unity at $z=0$, as expected by definition. However, notable differences arise at higher redshifts depending on the datasets employed.

When only low redshift probes (\textbf{LzP})  are considered, the reconstructed dark 
energy density exhibits a clear increasing trend with time. This behavior departs from the standard $\Lambda$CDM 
prediction, in which the dark energy density remains constant. The observed increase as time evolves therefore indicates a dynamical dark energy component induced by neutrino interactions. As discussed in  Sec. \ref{neutrino_beha}, this behavior is a generic consequence of the model and follows directly from the form of the neutrino dark energy coupling. On the other hand, the associated uncertainty band grows rapidly with redshift, reflecting the limited constraining power on the high $z$ behavior of the model.

In contrast, the inclusion of the CMB constraint through the acoustic angular scale $\theta_*$ noticeably alters the reconstruction. In this case, the evolution of $\rho_{\Lambda}(z)/\rho_{\Lambda}(0)$ is suppressed, remaining close to unity across the entire redshift range; however, it is 2$\sigma$ away from $\Lambda$CDM for $z \gtrsim 1$. Additionally, the uncertainty band becomes narrower, indicating an improvement in parameter constraints. This result highlights the critical role of early Universe information in constraining the late-time evolution of the dark energy parameter space.

\subsection{Cosmological inference for the two-neutrino model}\label{inference2}

\begin{figure}[h!]
\centering
\includegraphics[scale=0.5]{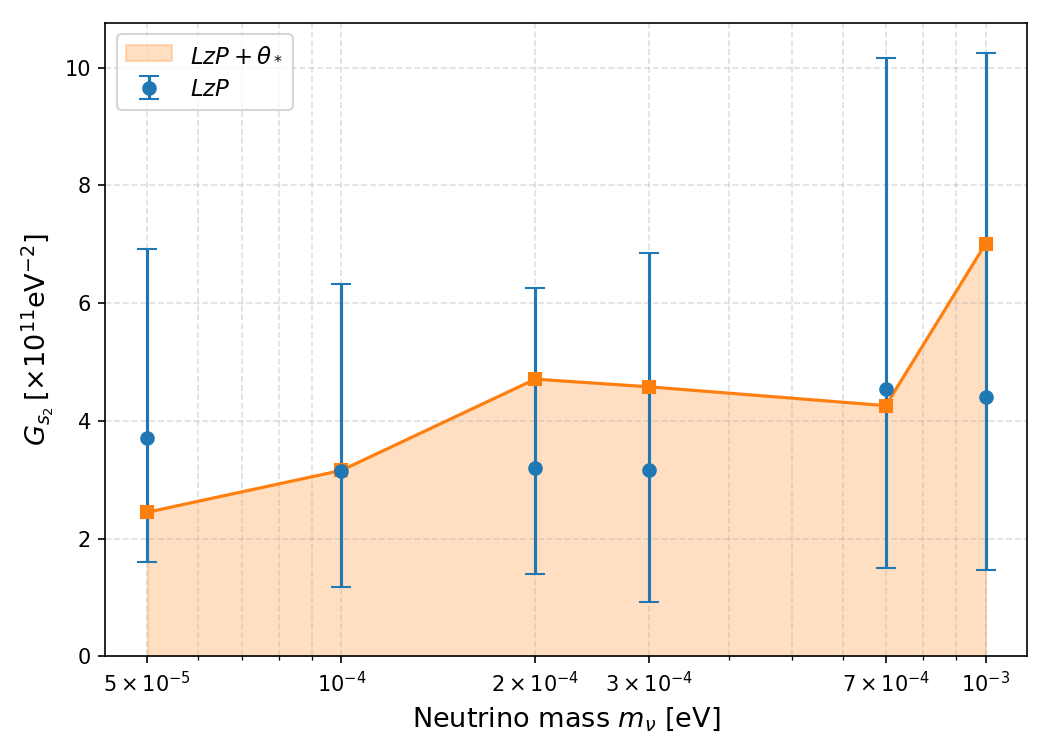}
\caption{Best fit values for $G_{s_2}$ (in blue) obtained from the following datasets:  Megamasers, Cosmic Chronometers, SNe Ia DES-Dovekie, and DESI BAO DR2 (\textbf{LzP}), whereas $G_{s_2}$ (in orange) uses the same before, but adding $\theta_*$ from Planck CMB (\textbf{LzP}+$\theta_*$).  }
\label{DD_all_chart_2}
\end{figure}

\begin{figure}[h!]
\centering
\includegraphics[scale=0.5]{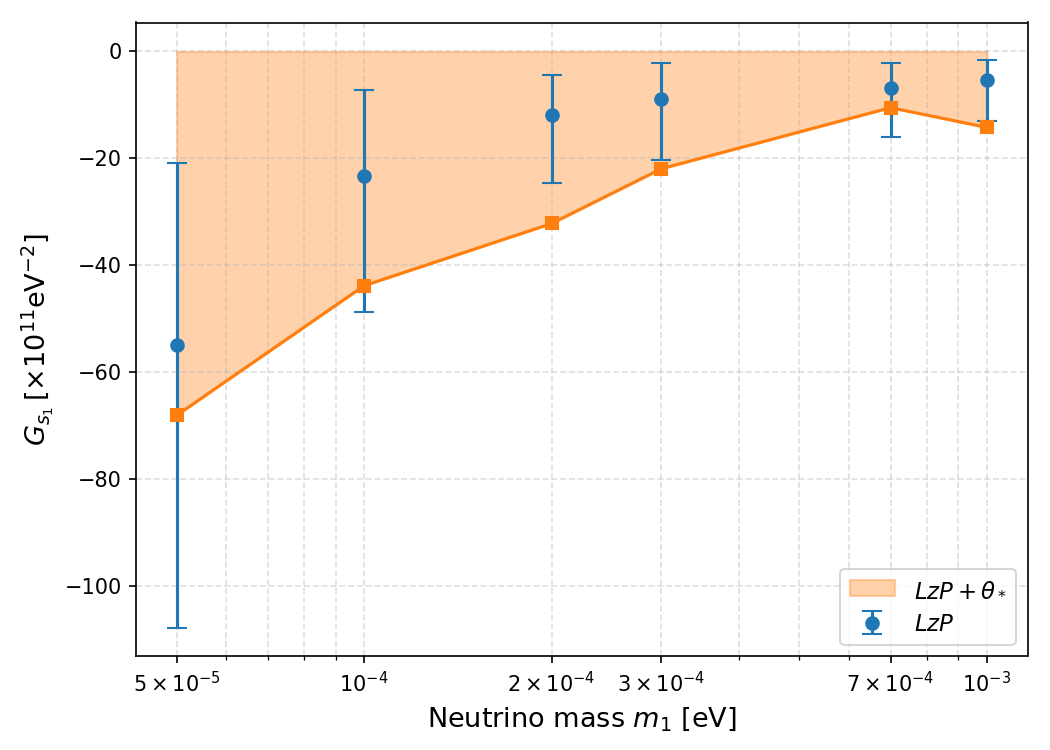}
\caption{Best fit values for $G_{s_1}$ (in blue) obtained from the following datasets:  Megamasers, Cosmic Chronometers, SNe Ia DES-Dovekie, and DESI BAO DR2 (\textbf{LzP}), whereas $G_{s_1}$ (in orange) uses the same before, but adding $\theta_*$ from Planck CMB (\textbf{LzP}+$\theta_*$).  }
\label{DD_all_chart_2_1}
\end{figure}

The dark energy dynamics is given by Eq. (\ref{rho_T12}), but at high temperatures, the dark energy is dominated by the heaviest neutrino that is relativistic, in this case $m_2$, hence given by Eq. \eqref{rho_ur}. If the effective coupling $G_{s_2}$ is positive one can have a beginning of $\rho_{\Lambda}$ from zero, but to avoid a divergence of $\rho_{\Lambda}$ for large $z$, one needs  a more fundamental microscopic realization, in which the interaction is dynamically suppressed in the early Universe and 
becomes relevant only at late times. On the other hand, we can choose $G_{s_1}<0$ to get a decreasing behavior of $\rho_{\Lambda}$ in the late universe as seen in the DESI results \cite{DESI:2025fii}. These constraints were obtained adopting the prior $\eta\in(-100,0.1)$, cf. Eq.~(\ref{eta}).
The $1\sigma$ confidence intervals of the effective couplings $G_{s_2}$ and $G_{s_1}$ for the two-neutrino model are reported in Table~\ref{G-val_2n_t}.  From the $G_{s_2}$ constraints, no clear monotonic dependence on the neutrino mass can be identified. The $G_{s_2}$ preferred values remain approximately constant, of order $(3$--$5)\times10^{11}\mathrm{eV}^{-2}$, throughout the mass range considered for the {\bf LzP} data. Although the largest neutrino masses tend to exhibit slightly higher central values, the associated uncertainties are substantial and largely overlap with those obtained for lower masses.  Therefore, within the sensitivity of the present data, we find no statistically significant evidence for a correlation between $G_{s_2}$ and $m_\nu$. This behavior is illustrated in Fig.~\ref{DD_all_chart_2}. A different trend is observed for $G_{s_1}$, whose preferred values exhibit a systematic dependence on the neutrino mass, as shown in Fig.~\ref{DD_all_chart_2_1}. In particular, the preferred values become progressively less negative as $m_1$ increases, changing from approximately $-5.5\times10^{12}\mathrm{eV}^{-2}$ at $m_1=5\times10^{-5}\mathrm{eV}$ to about $-6\times10^{11}\mathrm{eV}^{-2}$ at $m_1=10^{-3}\mathrm{eV}$ with \textbf{LzP} data. Although the associated uncertainties remain sizable, especially at low masses, the overall trend remains monotonic across the mass range considered. This behavior suggests a mild correlation between $G_{s_1}$ and $m_1$.

Additional insight is provided by the significance estimates $N_\sigma$
reported in Table~ \ref{G-val_2n_t}. In this case, we compute these as
$ N_\sigma = \Phi^{-1}(1-p/2)$, 
where $\Phi^{-1}$ is the inverse cumulative distribution function of the
standard normal distribution. This quantity gives the Gaussian-equivalent significance associated with the $p$-value obtained from
the likelihood-ratio statistic $\Delta\chi^2$, following standard
conventions in particle physics and cosmology \cite{PDG2022,Cowan1998}.

For the {\bf LzP} dataset, the preferred values of the interaction strength correspond to deviations from the non-interacting limit at the $1.8$--$2.1\sigma$ level, depending on the assumed neutrino mass. The strongest significance is found for $m_1 = 2.0 \times 10^{-4}\,\mathrm{eV}$, where the preference for a
nonzero interaction reaches about $2.1\sigma$. Although these values do not constitute decisive evidence, they indicate a mild preference for neutrino self-interactions when only the {\bf LzP} dataset is considered.
By contrast, for the combined {\bf LzP+$\theta_\ast$} data, the inferred constraints are consistent with the non-interacting limit within the reported confidence intervals. Therefore, no meaningful significance can be assigned to a nonzero interaction in that case, indicating that the preference observed with {\bf LzP} alone is not robust once the additional geometric information encoded in $\theta_\ast$ is included. Nevertheless, in our discussion we adopt the  value $m_1=5.0\times10^{-5}\,\mathrm{eV}$, since it provides the best qualitative reproduction of the dark energy evolution, comparing Fig. \ref{5e-5_3DD_rho_2} with the DESI DR2 analysis, \cite{DESI:2025fii} (see Fig. 9 in that document). In particular, this case exhibits a maximum in the dark-energy density at $z<1$, followed by a decrease toward its present-day value, closely matching the behavior inferred from the DESI reconstruction.

It is worth noting that the effective couplings $G_{s_j}$ are not fundamental parameters, but rather weighted sums of the microscopic interaction matrix elements $G_{s,ij}$. In a broad class of scalar-mediated interactions, the latter can be written as in Eq.~(\ref{effective_coup1}) \cite{Blinov:2019gcj,Berryman:2022hds}, where $\rm g_i$ denotes the coupling of the mediator field $\phi$ to the neutrino i-eigenstate. Since the sign of $G_{s,ij}$ is determined by the product $\rm g_i g_j$, individual matrix elements may naturally be either positive or negative depending on the relative signs of the underlying couplings. Consequently, the effective quantities $G_{s_j}$ may also acquire either sign after summation over the contributing species.

This sign asymmetry plays a crucial role in shaping the evolution of the effective dark energy density. Unlike the single-neutrino scenario, where the dark energy component evolves monotonically, the two-neutrino model is able to produce a transient enhancement of $\rho_\Lambda(z)$ at intermediate redshifts, followed by a decay toward the present epoch. 
As a result, the standard early-time cosmological evolution is preserved while the dark energy dynamics are modified at low redshift.

\begin{table*}[t]
\centering
\small
\begin{tabular}{ccccccc}
\toprule

& & \multicolumn{3}{c}{\textbf{LzP}}
& \multicolumn{2}{c}{\textbf{LzP+$\theta_*$}} \\
\cmidrule(lr){3-5}\cmidrule(lr){6-7}

$m_1\, [\mathrm{eV}]$ &
$m_2\, [\mathrm{eV}]$ &
$G_{s_2}\,$ &
$G_{s_1}\,$ & $N_{\sigma}$ & \,
$G_{s_2}\,$ &
$G_{s_1}\,$  \\
\midrule

$5.0\times10^{-5}$ & \, \,
$8.655\times10^{-3}$ &\, \,
$3.71^{+3.21}_{-2.11}$ &\, \,
$-54.87^{+33.91}_{-52.95}$ &\, \,
$1.8$ & \, \,
$<2.45$ &\, \,
$>-68.10$ \\

$1.0\times10^{-4}$ & \, \,
$8.655\times10^{-3}$ &\, \,
$3.16^{+3.17}_{-1.97}$ &\, \,
$-23.26^{+15.98}_{-25.59}$ &\, \,
$2.0$ & \, \,
$<3.16$ &\, \,
$>-43.84$ \\

$2.0\times10^{-4}$ & \, \,
$8.657\times10^{-3}$ & \, \,
$3.21^{+3.06}_{-1.81}$ & \, \,
$-11.84^{+7.43}_{-12.77}$ & \, \,
$2.1$ & \, \,
$<4.71$ & \, \,
$>-32.21$ \\

$3.0\times10^{-4}$ & \, \,
$8.660\times10^{-3}$ & \, \,
$3.17^{+3.69}_{-2.24}$ & \, \, 
$-8.91^{+6.77}_{-11.41}$ & \, \,
$2.0$ & \, \,
$<4.58$ & \, \,
$>-21.97$ \\

$7.0\times10^{-4}$ & \, \,
$8.683\times10^{-3}$ & \, \,
$4.54^{+5.63}_{-3.03}$ & \, \,
$-6.95^{+4.77}_{-9.02}$ & \, \,
$1.8$ & \, \,
$<4.26$ & \, \,
$>-10.54$ \\

$1.0\times10^{-3}$ & \, \,
$8.712\times10^{-3}$ & \, \,
$4.40^{+5.85}_{-2.93}$ & \, \,
$-5.41^{+3.74}_{-7.57}$ & \, \,
$1.8$ & \, \,
$<7.01$ &\, \,
$>-14.24$ \\

\bottomrule
\end{tabular}

\caption{Best-fit values at 68\% CL for $G_{s_2}$, $G_{s_1}$
$[\times 10^{11}\,\mathrm{eV}^{-2}]$ and the corresponding rough
significance estimates for the \textbf{LzP} analysis. $N_\sigma$ denotes the significance of the preference of the dynamical over the $\Lambda$CDM model.
The mass
$m_2$ is computed from Eq.~(\ref{Deltam2}) using
$\Delta m_{21}^2=7.49\times10^{-5}\,\mathrm{eV}^2$. }
\label{G-val_2n_t}
\end{table*}

For the smallest neutrino mass considered, $m_1=5\times10^{-5}\mathrm{eV}$, we obtain the $2\sigma$ constraint $G_{s_1}>-1.06\times10^{12}\mathrm{eV}^{-2}$ using the {\bf LzP} data. The addition of the CMB acoustic-scale measurement $\theta_\ast$ improves the limit to $G_{s_1}>-0.72\times10^{12}\mathrm{eV}^{-2}$.

\begin{figure}[h!]
\centering
\includegraphics[scale=0.8]{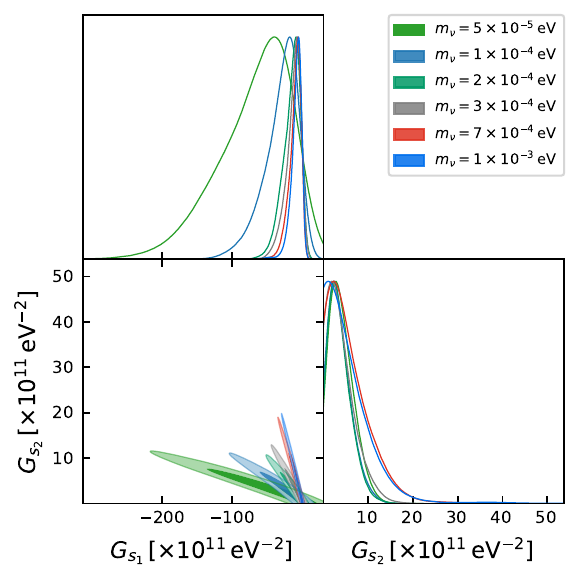}
\caption{Marginalized one- and two-dimensional posterior distributions for $G_{s_2}$ and $G_{s_1}$ in the two-neutrino model using Megamasers, Cosmic Chronometers, SNe Ia DES-Dovekie,  DESI BAO DR2, i. e., \textbf{LzP} . The no inclusion of $\theta_*$ leads to broader contours and enhanced degeneracies between the parameters.}
\label{G12}
\end{figure}

\begin{figure}[h!]
\centering
\includegraphics[scale=0.8]{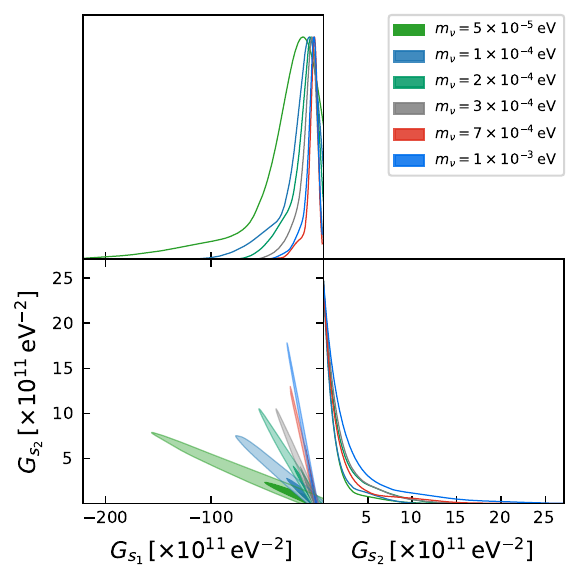}
\caption{Marginalized one and two dimensional posterior distributions for $G_{s_2}$ and $G_{s_1}$ in the two-neutrino model using Megamasers, Cosmic Chronometers, SNe Ia DES-Dovekie,  and  DESI BAO DR2, i. e., \textbf{LzP + $\theta_*$}. The contours show relatively tight constraints and correlations between the parameters.}
\label{G12sPl}
\end{figure}

Next we present the marginalized posteriors of $G_{s_2}$ and $G_{s_1}$ for different values of the lightest neutrino mass. Fig. ~\ref{G12} displays the results obtained with the \textbf{LzP} dataset. The posteriors exhibit well-defined maxima, indicating that both couplings are well constrained by the data. A correlation between $G_{s_2}$ and $G_{s_1}$ is observed, together with a dependence on the neutrino mass. Lower masses generally lead to broader posterior distributions, while larger masses result in tighter constraints. In contrast, Fig.~\ref{G12sPl} shows that including $\theta_*$ narrows the $G_{s_2}$ posterior distribution, and similar degeneracies are observed. 

\begin{figure}[h!]
\centering
\includegraphics[scale=0.38]{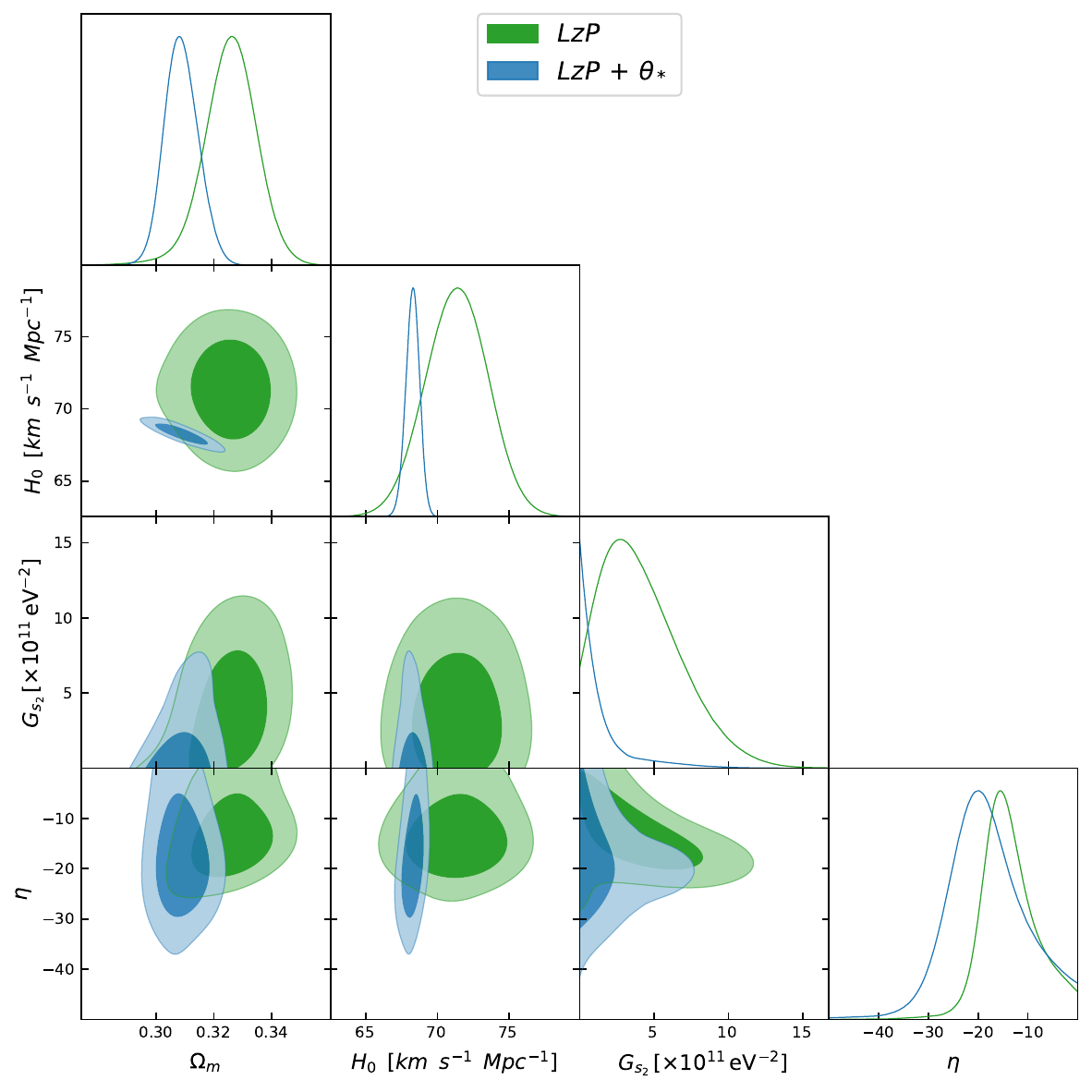}
\caption{\normalsize Comparison of the marginalized posterior distributions for $m_\nu = 5 \times 10^{-5}\,\mathrm{eV}$, which corresponds to the case that exhibits the largest deviation from $\Lambda$CDM among those studied. The analysis uses \textbf{LzP} dataset (green), and the same dataset combination including $\theta_*$ (blue). The shaded contours correspond to the $1\sigma$ and $2\sigma$ confidence levels.}
\label{DD_all_2n}
\end{figure}

\begin{figure}[h!]
\centering
\includegraphics[scale=0.4]{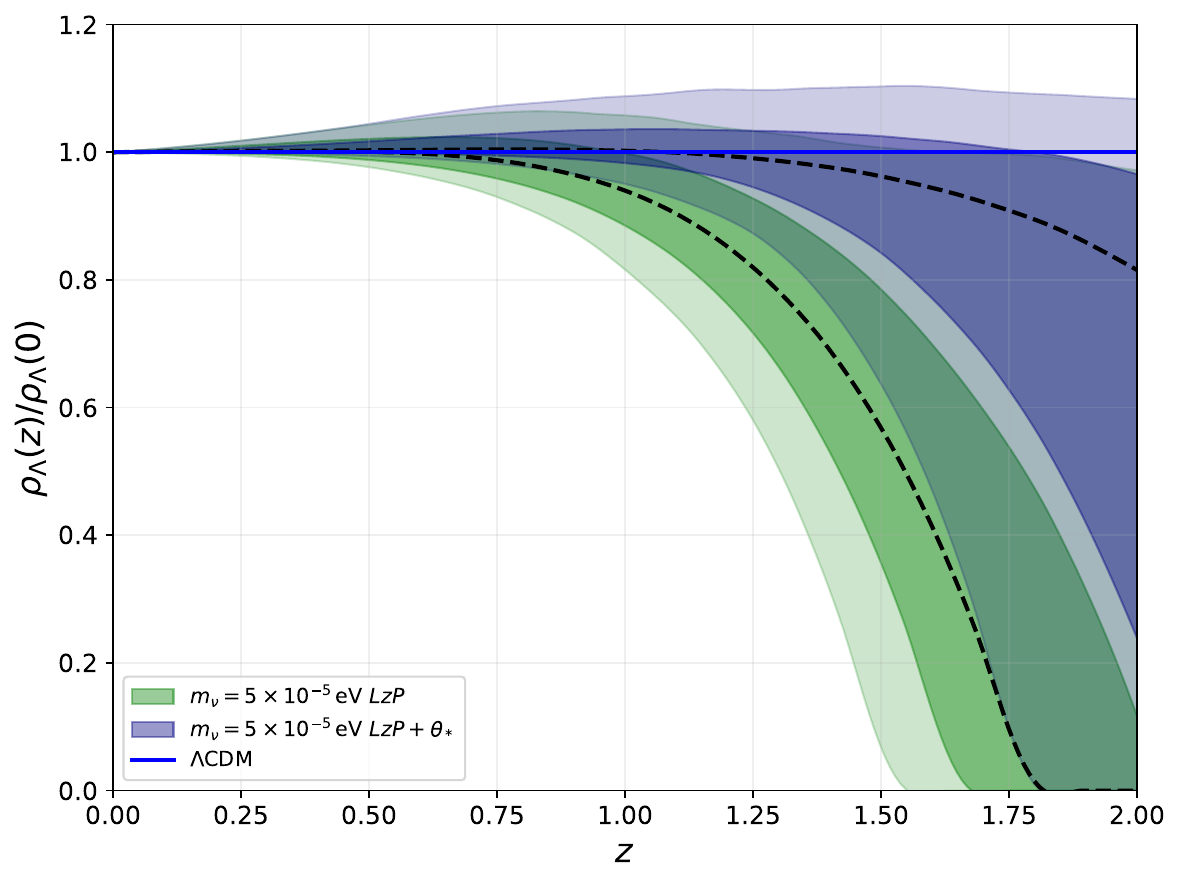}
\caption{
Reconstructed evolution of $\rho_{\Lambda}(z)/\rho_{\Lambda}(0)$ in the two-neutrino model for $m_{\nu}=5\times10^{-5}\mathrm{eV}$. Green and blue regions correspond to the constraints obtained from the \textbf{LzP}  and \textbf{LzP+$\theta_\ast$} datasets, respectively. Dark and light shaded bands indicate the $68\%$ and $95\%$ confidence intervals, while dashed lines show the median reconstruction. The horizontal blue line denotes the $\Lambda$CDM prediction. The inclusion of the $\theta_\ast$ measurement shifts the reconstruction toward $\Lambda$CDM and increases the uncertainty at higher redshifts. }
\label{5e-5_3DD_rho_2}
\end{figure}

In Fig. \ref{DD_all_2n}, the posterior distributions of $G_{s_2}$ together with $\Omega_{m}$, $H_0$, and $\eta$ for $m_\nu = 5\times10^{-5}\mathrm{eV}$, are compared, analyzing both with and without $\theta_*$. 

In agreement with Figs.~\ref{G12} and \ref{G12sPl}, the inclusion of $\theta_*$ shifts $G_{s_2}$ toward smaller values and introduces clear degeneracies, particularly with $\eta$ and, to a lesser extent, with $H_0$. These correlations explain the degradation of the constraints observed when $\theta_*$ is included.

On the other hand, Fig. \ref{5e-5_3DD_rho_2} shows the evolution of $\rho_\Lambda(z)/\rho_\Lambda(0)$ for $m_\nu = 5\times10^{-5} \, \mathrm{eV}$, comparing the analyses with and without $\theta_*$. Consistent with the previous results, the inclusion of $\theta_*$ leads to a broader allowed region and a milder evolution of $\rho_\Lambda(z)$, keeping it closer to a constant behavior. In contrast, without $\theta_*$, the deviation from $\Lambda$CDM becomes more pronounced at higher redshifts.

\section{Phenomenological implications} \label{discussion}
The preferred values of the interaction parameters remain of comparable magnitude across the entire mass range considered, with substantial overlap among their confidence intervals. Therefore, 
the cosmological analysis performed in this work constrains the effective scalar-neutrino coupling parameter $G_{s_j}$ to values of order 
\begin{equation}
    G_{s_j} \sim 10^{12}\ \mathrm{eV}^{-2} .
\end{equation}

The preferred values of $G_{s_j}$ obtained in our analysis are substantially larger than the effective interaction strengths typically considered in studies of secret neutrino interactions and invisible neutrino decays, as reviewed in Ref.~\cite{EscuderoRelaxing}. However, a direct numerical comparison should be made with caution. In the present framework, $G_{s_j}$ enters through the energy-transfer current $J_0$, Eq.~(\ref{J_R}), and therefore characterizes the coupling responsible for the exchange of energy between neutrinos and the dark energy sector at the cosmological level. By contrast, the couplings compiled in Ref.~\cite{EscuderoRelaxing} are generally associated with microscopic particle interactions, commonly parameterized in terms of dimensionless couplings and mediator masses. Consequently, although both descriptions involve neutrino interactions, the corresponding coupling strengths are defined in different physical contexts and are not directly comparable on a one-to-one basis.

Using the definition of  $G_{s_j}$ introduced in Eq. \eqref{effective_coup},
one can estimate the characteristic mass scale of the scalar mediator. Assuming a perturbative neutrino coupling of order unity, $\rm g_i \sim \mathcal{O}(1)$, the inferred mediator mass becomes
\begin{equation}
    m_\phi \sim G_{s_j}^{-1/2}.
\end{equation}
For the typical values obtained in our analysis,
we find $m_\phi \sim 10^{-6}\ \mathrm{eV}$. Therefore, the mediator associated with the neutrino interaction lies naturally in the ultralight regime.

Interestingly, this mass scale overlaps with parameter regions commonly discussed for axion-like particles and ultralight scalar mediators appearing in extensions of the Standard Model. Recent reviews and phenomenological analyses, such as \cite{OHare:2024nmr}, indicate that viable ultralight bosons may span masses between 
\begin{equation}
10^{-33}\ \mathrm{eV}
\lesssim
m_\phi
\lesssim
1\ \mathrm{eV},
\end{equation}
depending on the cosmological and astrophysical scenario considered. In particular, axion-like particles motivated by string compactification frequently populate the range $10^{-12}\text{--}10^{-3}\ \mathrm{eV}$, which includes the mass scale inferred from our cosmological constraints. However, the mediator considered here is not necessarily the QCD axion. Rather, it should be viewed as a generic light scalar coupled to neutrinos. The framework remains strongly inspired by particle-physics constructions, being characterized by a small set of physical parameters, namely the neutrino masses $m_j$ and the scalar-neutrino couplings $G_{s_j}$. In this context, the scalar field acts as an effective mediator that induces temperature-dependent neutrino mass corrections, leading to a dynamical dark energy component. While the resulting cosmological phenomenology differs from that of conventional axion dark matter scenarios, the underlying structure retains the simplicity and motivation of particle-physics models involving light scalar mediators.

Moreover, the inferred mediator mass scale,
$m_\phi \sim 10^{-6}\,\mathrm{eV}$,
is remarkably close to the scales explored in several models of
secret neutrino interactions and ultralight scalar cosmologies
\cite{EscuderoRelaxing,Marsh2016,Ferreira2021}.
Such masses are sufficiently small to affect cosmological evolution
while remaining difficult to probe in terrestrial experiments.
Consequently, cosmological observations provide one of the most
sensitive avenues for testing this class of scenarios
\cite{EscuderoRelaxing,CyrRacine2014,Oldengott2017}.

\section{Conclusions} \label{conclusions}

We have explored a dark energy scenario in which finite-temperature
corrections to neutrino masses, generated by interactions mediated by a
light scalar field, contribute to an effective dark energy component
within the framework of Unimodular Gravity. In this construction, the
non-conservation current associated with the neutrino sector modifies an
otherwise constant vacuum-energy contribution, leading to a dynamical
dark energy density. 

We investigated two realizations of this scenario. In the one-neutrino model, the dark energy density evolves dynamically and increases monotonically toward the present epoch. The two-neutrino realization exhibits a richer phenomenology: depending on the relative signs and strengths of the effective couplings, the dark energy density may reach a maximum at low redshift, typically around ($z\lesssim1$), before decreasing toward the present epoch. As illustrated in Fig.~\ref{5e-5_3DD_rho_2}, this turnover occurs within the redshift range currently probed by late-time cosmological observations. Therefore, while both scenarios predict departures from a constant dark energy density, only the two-neutrino model naturally allows for a low-redshift maximum followed by a subsequent decline.

Using Megamasers, Cosmic Chronometers, DES-Dovekie SNe Ia, and DESI DR2
BAO measurements, dubbed \textbf{LzP} dataset, we constrained the effective interaction strength
$G_s$ for values of the lightest neutrino mass in the range 
$0.05\,{\rm meV} \le m_1 \le 1\,{\rm meV}$.
For the one-neutrino scenario, 
the preferred coupling values lie in the range $G_{s_1} \sim (2 \, \text{--} \, 25)\times10^{11}\,\mathrm{eV}^{-2}$ at 68\% CL, depending on the assumed neutrino mass. In particular, larger values of $G_{s_1}$ are favored for lighter neutrino masses, whereas heavier neutrinos require smaller interaction strengths. For the two-neutrino scenario, 
the inferred couplings are $G_{s_2} \sim 4 \times 10^{11}$ eV$^{-2}$ and $G_{s_1} \sim (-55 \, ,   \,-5) \times 10^{11}$ eV$^{-2}$ at 68\% CL over the explored neutrino-mass range. Assuming an underlying neutrino coupling of order unity,
the characteristic interaction scale corresponds to an
ultralight mediator with mass $m_\phi \sim 10^{-6}\mathrm{eV}$.

We further examined the impact of
including the Planck distance-prior information through the acoustic
scale $\theta_*$. We find that the addition of CMB information reduces the allowed parameter space.  
For the one-neutrino model, only the two smallest neutrino
masses retain a preference for non-zero interactions,
whereas in larger masses 
$G_{s_1}$ is consistent with zero at $1\sigma$.
For the two-neutrino model, the inclusion of Planck information
weakens the preference for non-zero interactions,
yielding only upper limits on both coupling parameters across the
explored neutrino-mass range. 
These constraints, in turn, generally drive the reconstructed dark energy
evolution closer to that of a constant dark energy component, highlighting 
the importance of combining 
early- and late-time observations when testing dynamical dark energy scenarios.

Regarding the statistical significance of departures from the $\Lambda$CDM limit ($G_{s_1}=0$), we find that, among the neutrino masses considered, the one-neutrino case exhibits an approximately $2\sigma$ departure, when the \textbf{LzP} dataset is employed. For the \textbf{LzP+$\theta_\ast$} combination, this departure is reduced to $1.5$--$1.6\sigma$ for masses $m_1 \leq 0.1 \, \mathrm{meV}$, while for larger masses only upper limits are obtained. This behavior reflects the tighter constraints induced by the inclusion of CMB acoustic-scale information and the resulting reduction of parameter degeneracies. The two-neutrino case exhibits a similar trend, with an approximately $2\sigma$ departure from the $\Lambda$CDM limit when the \textbf{LzP} dataset is employed, while the constraints for the \textbf{LzP+$\theta_\ast$} dataset allow negative values of $G_{s_1}$ only down to a finite lower bound.

Although the inferred dark energy evolution depends on the neutrino-mass
assumptions and on the dataset combination employed, the analysis shows
that non-standard neutrino interactions can generate cosmologically
relevant dark energy dynamics while remaining compatible with current
observational constraints. In particular, the model naturally recovers
the $\Lambda$CDM limit when the interaction contribution becomes
negligible, providing a consistent extension rather than a disconnected
alternative to the standard cosmological model.

The present analysis should be understood as a late-time effective 
description of a neutrino-induced dark energy component. Within this 
effective treatment, the neutrino--scalar interaction is not assumed to 
operate with the same form throughout the entire thermal history of the 
Universe. Indeed, a formal extrapolation of the late-time solution to 
sufficiently high temperatures can extend the model beyond its intended 
regime of validity. We therefore implement a transition between the 
relativistic and nonrelativistic regimes of the neutrino fluid and 
apply each solution within the corresponding physical domain. This 
matching prescription provides a controlled phenomenological 
description of the late-time dynamics probed by the datasets 
considered here. A more fundamental microscopic realization, in which 
the interaction is dynamically suppressed in the early Universe and 
becomes relevant only at late times, would provide a natural completion 
of this framework and represents an important direction for future 
work.

Future work should also include a full treatment of cosmological perturbations, CMB anisotropies, and large-scale-structure observables. Such analyses will be essential to further assess the observational viability of the model. More importantly, they will determine whether the low-redshift evolution of the dark energy density predicted in the two-neutrino scenario, including the possibility of a maximum followed by a decline toward the present epoch, can provide a viable explanation for the emerging indications of dynamical dark energy reported by recent observations.

\section{Acknowledgments}
This work was supported by SECIHTI (CONAHCyT) project CBF2023-2024-589 (JLCC) and through the postdoctoral scholarship, CVU No. 552250 (VAGO). M. L. H. was supported by SECIHTI grant 806098.

\newpage
 \bibliographystyle{apsrev4-2}
 \bibliography{references}

\begin{thebibliography}{82}%
\makeatletter
\providecommand \@ifxundefined [1]{%
 \@ifx{#1\undefined}
}%
\providecommand \@ifnum [1]{%
 \ifnum #1\expandafter \@firstoftwo
 \else \expandafter \@secondoftwo
 \fi
}%
\providecommand \@ifx [1]{%
 \ifx #1\expandafter \@firstoftwo
 \else \expandafter \@secondoftwo
 \fi
}%
\providecommand \natexlab [1]{#1}%
\providecommand \enquote  [1]{``#1''}%
\providecommand \bibnamefont  [1]{#1}%
\providecommand \bibfnamefont [1]{#1}%
\providecommand \citenamefont [1]{#1}%
\providecommand \href@noop [0]{\@secondoftwo}%
\providecommand \href [0]{\begingroup \@sanitize@url \@href}%
\providecommand \@href[1]{\@@startlink{#1}\@@href}%
\providecommand \@@href[1]{\endgroup#1\@@endlink}%
\providecommand \@sanitize@url [0]{\catcode `\\12\catcode `\$12\catcode `\&12\catcode `\#12\catcode `\^12\catcode `\_12\catcode `\%12\relax}%
\providecommand \@@startlink[1]{}%
\providecommand \@@endlink[0]{}%
\providecommand \url  [0]{\begingroup\@sanitize@url \@url }%
\providecommand \@url [1]{\endgroup\@href {#1}{\urlprefix }}%
\providecommand \urlprefix  [0]{URL }%
\providecommand \Eprint [0]{\href }%
\providecommand \doibase [0]{https://doi.org/}%
\providecommand \selectlanguage [0]{\@gobble}%
\providecommand \bibinfo  [0]{\@secondoftwo}%
\providecommand \bibfield  [0]{\@secondoftwo}%
\providecommand \translation [1]{[#1]}%
\providecommand \BibitemOpen [0]{}%
\providecommand \bibitemStop [0]{}%
\providecommand \bibitemNoStop [0]{.\EOS\space}%
\providecommand \EOS [0]{\spacefactor3000\relax}%
\providecommand \BibitemShut  [1]{\csname bibitem#1\endcsname}%
\let\auto@bib@innerbib\@empty
\bibitem [{\citenamefont {Adame}\ \emph {et~al.}(2025{\natexlab{a}})\citenamefont {Adame} \emph {et~al.}}]{DESI:2024mwx}%
  \BibitemOpen
  \bibfield  {author} {\bibinfo {author} {\bibfnamefont {A.~G.}\ \bibnamefont {Adame}} \emph {et~al.} (\bibinfo {collaboration} {DESI}),\ }\href {https://doi.org/10.1088/1475-7516/2025/02/021} {\bibfield  {journal} {\bibinfo  {journal} {JCAP}\ }\textbf {\bibinfo {volume} {02}}\bibfield  {number} {\bibinfo  {number} { (02)},\ \bibinfo {pages} {021}},\ }\Eprint {https://arxiv.org/abs/2404.03002} {arXiv:2404.03002 [astro-ph.CO]} \BibitemShut {NoStop}%
\bibitem [{\citenamefont {Abdul~Karim}\ \emph {et~al.}(2025)\citenamefont {Abdul~Karim} \emph {et~al.}}]{DESI:2025zgx}%
  \BibitemOpen
  \bibfield  {author} {\bibinfo {author} {\bibfnamefont {M.}~\bibnamefont {Abdul~Karim}} \emph {et~al.} (\bibinfo {collaboration} {DESI}),\ }\href {https://doi.org/10.1103/tr6y-kpc6} {\bibfield  {journal} {\bibinfo  {journal} {Phys. Rev. D}\ }\textbf {\bibinfo {volume} {112}},\ \bibinfo {pages} {083515} (\bibinfo {year} {2025})},\ \Eprint {https://arxiv.org/abs/2503.14738} {arXiv:2503.14738 [astro-ph.CO]} \BibitemShut {NoStop}%
\bibitem [{\citenamefont {Aghanim}\ \emph {et~al.}(2020)\citenamefont {Aghanim} \emph {et~al.}}]{Planck:2019nip}%
  \BibitemOpen
  \bibfield  {author} {\bibinfo {author} {\bibfnamefont {N.}~\bibnamefont {Aghanim}} \emph {et~al.} (\bibinfo {collaboration} {Planck}),\ }\href {https://doi.org/10.1051/0004-6361/201936386} {\bibfield  {journal} {\bibinfo  {journal} {Astron. Astrophys.}\ }\textbf {\bibinfo {volume} {641}},\ \bibinfo {pages} {A5} (\bibinfo {year} {2020})},\ \Eprint {https://arxiv.org/abs/1907.12875} {arXiv:1907.12875 [astro-ph.CO]} \BibitemShut {NoStop}%
\bibitem [{\citenamefont {Scolnic}\ \emph {et~al.}(2022)\citenamefont {Scolnic} \emph {et~al.}}]{Scolnic:2021amr}%
  \BibitemOpen
  \bibfield  {author} {\bibinfo {author} {\bibfnamefont {D.}~\bibnamefont {Scolnic}} \emph {et~al.},\ }\href {https://doi.org/10.3847/1538-4357/ac8b7a} {\bibfield  {journal} {\bibinfo  {journal} {Astrophys. J.}\ }\textbf {\bibinfo {volume} {938}},\ \bibinfo {pages} {113} (\bibinfo {year} {2022})},\ \Eprint {https://arxiv.org/abs/2112.03863} {arXiv:2112.03863 [astro-ph.CO]} \BibitemShut {NoStop}%
\bibitem [{\citenamefont {Brout}\ \emph {et~al.}(2022)\citenamefont {Brout} \emph {et~al.}}]{Brout:2022vxf}%
  \BibitemOpen
  \bibfield  {author} {\bibinfo {author} {\bibfnamefont {D.}~\bibnamefont {Brout}} \emph {et~al.},\ }\href {https://doi.org/10.3847/1538-4357/ac8e04} {\bibfield  {journal} {\bibinfo  {journal} {Astrophys. J.}\ }\textbf {\bibinfo {volume} {938}},\ \bibinfo {pages} {110} (\bibinfo {year} {2022})},\ \Eprint {https://arxiv.org/abs/2202.04077} {arXiv:2202.04077 [astro-ph.CO]} \BibitemShut {NoStop}%
\bibitem [{\citenamefont {Rubin}\ \emph {et~al.}(2025)\citenamefont {Rubin} \emph {et~al.}}]{Rubin:2023jdq}%
  \BibitemOpen
  \bibfield  {author} {\bibinfo {author} {\bibfnamefont {D.}~\bibnamefont {Rubin}} \emph {et~al.},\ }\href {https://doi.org/10.3847/1538-4357/adc0a5} {\bibfield  {journal} {\bibinfo  {journal} {Astrophys. J.}\ }\textbf {\bibinfo {volume} {986}},\ \bibinfo {pages} {231} (\bibinfo {year} {2025})},\ \Eprint {https://arxiv.org/abs/2311.12098} {arXiv:2311.12098 [astro-ph.CO]} \BibitemShut {NoStop}%
\bibitem [{\citenamefont {Abbott}\ \emph {et~al.}(2024{\natexlab{a}})\citenamefont {Abbott} \emph {et~al.}}]{DES:2024jxu}%
  \BibitemOpen
  \bibfield  {author} {\bibinfo {author} {\bibfnamefont {T.~M.~C.}\ \bibnamefont {Abbott}} \emph {et~al.} (\bibinfo {collaboration} {DES}),\ }\href {https://doi.org/10.3847/2041-8213/ad6f9f} {\bibfield  {journal} {\bibinfo  {journal} {Astrophys. J. Lett.}\ }\textbf {\bibinfo {volume} {973}},\ \bibinfo {pages} {L14} (\bibinfo {year} {2024}{\natexlab{a}})},\ \Eprint {https://arxiv.org/abs/2401.02929} {arXiv:2401.02929 [astro-ph.CO]} \BibitemShut {NoStop}%
\bibitem [{\citenamefont {Calderon}\ \emph {et~al.}(2024)\citenamefont {Calderon} \emph {et~al.}}]{DESI:2024aqx}%
  \BibitemOpen
  \bibfield  {author} {\bibinfo {author} {\bibfnamefont {R.}~\bibnamefont {Calderon}} \emph {et~al.} (\bibinfo {collaboration} {DESI}),\ }\href {https://doi.org/10.1088/1475-7516/2024/10/048} {\bibfield  {journal} {\bibinfo  {journal} {JCAP}\ }\textbf {\bibinfo {volume} {10}}\bibfield  {number} {\bibinfo  {number} { (10)},\ \bibinfo {pages} {048}},\ }\Eprint {https://arxiv.org/abs/2405.04216} {arXiv:2405.04216 [astro-ph.CO]} \BibitemShut {NoStop}%
\bibitem [{\citenamefont {Lodha}\ \emph {et~al.}(2025{\natexlab{a}})\citenamefont {Lodha} \emph {et~al.}}]{DESI:2024kob}%
  \BibitemOpen
  \bibfield  {author} {\bibinfo {author} {\bibfnamefont {K.}~\bibnamefont {Lodha}} \emph {et~al.} (\bibinfo {collaboration} {DESI}),\ }\href {https://doi.org/10.1103/PhysRevD.111.023532} {\bibfield  {journal} {\bibinfo  {journal} {Phys. Rev. D}\ }\textbf {\bibinfo {volume} {111}},\ \bibinfo {pages} {023532} (\bibinfo {year} {2025}{\natexlab{a}})},\ \Eprint {https://arxiv.org/abs/2405.13588} {arXiv:2405.13588 [astro-ph.CO]} \BibitemShut {NoStop}%
\bibitem [{\citenamefont {Lodha}\ \emph {et~al.}(2025{\natexlab{b}})\citenamefont {Lodha} \emph {et~al.}}]{DESI:2025fii}%
  \BibitemOpen
  \bibfield  {author} {\bibinfo {author} {\bibfnamefont {K.}~\bibnamefont {Lodha}} \emph {et~al.} (\bibinfo {collaboration} {DESI}),\ }\href {https://doi.org/10.1103/w4c6-1r5j} {\bibfield  {journal} {\bibinfo  {journal} {Phys. Rev. D}\ }\textbf {\bibinfo {volume} {112}},\ \bibinfo {pages} {083511} (\bibinfo {year} {2025}{\natexlab{b}})},\ \Eprint {https://arxiv.org/abs/2503.14743} {arXiv:2503.14743 [astro-ph.CO]} \BibitemShut {NoStop}%
\bibitem [{\citenamefont {Adame}\ \emph {et~al.}(2025{\natexlab{b}})\citenamefont {Adame} \emph {et~al.}}]{DESI:2024hhd}%
  \BibitemOpen
  \bibfield  {author} {\bibinfo {author} {\bibfnamefont {A.~G.}\ \bibnamefont {Adame}} \emph {et~al.} (\bibinfo {collaboration} {DESI}),\ }\href {https://doi.org/10.1088/1475-7516/2025/07/028} {\bibfield  {journal} {\bibinfo  {journal} {JCAP}\ }\textbf {\bibinfo {volume} {07}}\bibfield  {number} {\bibinfo  {number} { (07)},\ \bibinfo {pages} {028}},\ }\Eprint {https://arxiv.org/abs/2411.12022} {arXiv:2411.12022 [astro-ph.CO]} \BibitemShut {NoStop}%
\bibitem [{\citenamefont {Esteban}\ \emph {et~al.}(2024)\citenamefont {Esteban}, \citenamefont {Gonzalez-Garcia}, \citenamefont {Maltoni}, \citenamefont {Martinez-Soler}, \citenamefont {Pinheiro},\ and\ \citenamefont {Schwetz}}]{Esteban:2024eli}%
  \BibitemOpen
  \bibfield  {author} {\bibinfo {author} {\bibfnamefont {I.}~\bibnamefont {Esteban}}, \bibinfo {author} {\bibfnamefont {M.~C.}\ \bibnamefont {Gonzalez-Garcia}}, \bibinfo {author} {\bibfnamefont {M.}~\bibnamefont {Maltoni}}, \bibinfo {author} {\bibfnamefont {I.}~\bibnamefont {Martinez-Soler}}, \bibinfo {author} {\bibfnamefont {J.~P.}\ \bibnamefont {Pinheiro}},\ and\ \bibinfo {author} {\bibfnamefont {T.}~\bibnamefont {Schwetz}},\ }\href {https://doi.org/10.1007/JHEP12(2024)216} {\bibfield  {journal} {\bibinfo  {journal} {JHEP}\ }\textbf {\bibinfo {volume} {12}}\bibfield  {number} {\bibinfo  {number} { (12)},\ \bibinfo {pages} {216}},\ }\Eprint {https://arxiv.org/abs/2410.05380} {arXiv:2410.05380 [hep-ph]} \BibitemShut {NoStop}%
\bibitem [{\citenamefont {Camphuis}\ \emph {et~al.}(2026)\citenamefont {Camphuis} \emph {et~al.}}]{SPT-3G:2025bzu}%
  \BibitemOpen
  \bibfield  {author} {\bibinfo {author} {\bibfnamefont {E.}~\bibnamefont {Camphuis}} \emph {et~al.} (\bibinfo {collaboration} {SPT-3G}),\ }\href {https://doi.org/10.1103/7wt3-9v2y} {\bibfield  {journal} {\bibinfo  {journal} {Phys. Rev. D}\ }\textbf {\bibinfo {volume} {113}},\ \bibinfo {pages} {083504} (\bibinfo {year} {2026})},\ \Eprint {https://arxiv.org/abs/2506.20707} {arXiv:2506.20707 [astro-ph.CO]} \BibitemShut {NoStop}%
\bibitem [{\citenamefont {Elbers}\ \emph {et~al.}(2025)\citenamefont {Elbers} \emph {et~al.}}]{Elbers:2025vlz}%
  \BibitemOpen
  \bibfield  {author} {\bibinfo {author} {\bibfnamefont {W.}~\bibnamefont {Elbers}} \emph {et~al.},\ }\href {https://doi.org/10.1103/w9pk-xsk7} {\bibfield  {journal} {\bibinfo  {journal} {Phys. Rev. D}\ }\textbf {\bibinfo {volume} {112}},\ \bibinfo {pages} {083513} (\bibinfo {year} {2025})},\ \Eprint {https://arxiv.org/abs/2503.14744} {arXiv:2503.14744 [astro-ph.CO]} \BibitemShut {NoStop}%
\bibitem [{\citenamefont {Pulido-Hern{\'a}ndez}\ and\ \citenamefont {Cervantes-Cota}(2026)}]{Pulido-Hernandez:2026hcs}%
  \BibitemOpen
  \bibfield  {author} {\bibinfo {author} {\bibfnamefont {H.}~\bibnamefont {Pulido-Hern{\'a}ndez}}\ and\ \bibinfo {author} {\bibfnamefont {J.~L.}\ \bibnamefont {Cervantes-Cota}},\ }\href@noop {} {\bibfield  {journal} {\bibinfo  {journal} {arXiv e-prints}\ } (\bibinfo {year} {2026})},\ \Eprint {https://arxiv.org/abs/2603.13208} {arXiv:2603.13208 [astro-ph.CO]} \BibitemShut {NoStop}%
\bibitem [{\citenamefont {Sailer}\ \emph {et~al.}(2026)\citenamefont {Sailer}, \citenamefont {Farren}, \citenamefont {Ferraro},\ and\ \citenamefont {White}}]{Sailer:2025lxj}%
  \BibitemOpen
  \bibfield  {author} {\bibinfo {author} {\bibfnamefont {N.}~\bibnamefont {Sailer}}, \bibinfo {author} {\bibfnamefont {G.~S.}\ \bibnamefont {Farren}}, \bibinfo {author} {\bibfnamefont {S.}~\bibnamefont {Ferraro}},\ and\ \bibinfo {author} {\bibfnamefont {M.}~\bibnamefont {White}},\ }\href {https://doi.org/10.1103/6r54-8lv4} {\bibfield  {journal} {\bibinfo  {journal} {Phys. Rev. Lett.}\ }\textbf {\bibinfo {volume} {136}},\ \bibinfo {pages} {081002} (\bibinfo {year} {2026})},\ \Eprint {https://arxiv.org/abs/2504.16932} {arXiv:2504.16932 [astro-ph.CO]} \BibitemShut {NoStop}%
\bibitem [{\citenamefont {Aguilar}\ \emph {et~al.}(2001)\citenamefont {Aguilar} \emph {et~al.}}]{LSND2001}%
  \BibitemOpen
  \bibfield  {author} {\bibinfo {author} {\bibfnamefont {A.}~\bibnamefont {Aguilar}} \emph {et~al.} (\bibinfo {collaboration} {LSND Collaboration}),\ }\href {https://doi.org/10.1103/PhysRevD.64.112007} {\bibfield  {journal} {\bibinfo  {journal} {Phys. Rev. D}\ }\textbf {\bibinfo {volume} {64}},\ \bibinfo {pages} {112007} (\bibinfo {year} {2001})}\BibitemShut {NoStop}%
\bibitem [{\citenamefont {Aguilar-Arevalo}\ \emph {et~al.}(2018)\citenamefont {Aguilar-Arevalo}, \citenamefont {Brown}, \citenamefont {Bugel}, \citenamefont {Cheng}, \citenamefont {Conrad}, \citenamefont {Cooper} \emph {et~al.}}]{MiniBoone2018}%
  \BibitemOpen
  \bibfield  {author} {\bibinfo {author} {\bibfnamefont {A.~A.}\ \bibnamefont {Aguilar-Arevalo}}, \bibinfo {author} {\bibfnamefont {B.~C.}\ \bibnamefont {Brown}}, \bibinfo {author} {\bibfnamefont {L.}~\bibnamefont {Bugel}}, \bibinfo {author} {\bibfnamefont {G.}~\bibnamefont {Cheng}}, \bibinfo {author} {\bibfnamefont {J.~M.}\ \bibnamefont {Conrad}}, \bibinfo {author} {\bibfnamefont {R.~L.}\ \bibnamefont {Cooper}}, \emph {et~al.} (\bibinfo {collaboration} {MiniBooNE Collaboration}),\ }\href {https://doi.org/10.1103/PhysRevLett.121.221801} {\bibfield  {journal} {\bibinfo  {journal} {Phys. Rev. Lett.}\ }\textbf {\bibinfo {volume} {121}},\ \bibinfo {pages} {221801} (\bibinfo {year} {2018})}\BibitemShut {NoStop}%
\bibitem [{\citenamefont {Palomares-Ruiz}\ \emph {et~al.}(2005)\citenamefont {Palomares-Ruiz}, \citenamefont {Pascoli},\ and\ \citenamefont {Schwetz}}]{Palomares2005}%
  \BibitemOpen
  \bibfield  {author} {\bibinfo {author} {\bibfnamefont {S.}~\bibnamefont {Palomares-Ruiz}}, \bibinfo {author} {\bibfnamefont {S.}~\bibnamefont {Pascoli}},\ and\ \bibinfo {author} {\bibfnamefont {T.}~\bibnamefont {Schwetz}},\ }\href {https://doi.org/10.1088/1126-6708/2005/09/048} {\bibfield  {journal} {\bibinfo  {journal} {Journal of High Energy Physics}\ }\textbf {\bibinfo {volume} {2005}},\ \bibinfo {pages} {048} (\bibinfo {year} {2005})}\BibitemShut {NoStop}%
\bibitem [{\citenamefont {Bell}\ \emph {et~al.}(2006)\citenamefont {Bell}, \citenamefont {Pierpaoli},\ and\ \citenamefont {Sigurdson}}]{Bell2006}%
  \BibitemOpen
  \bibfield  {author} {\bibinfo {author} {\bibfnamefont {N.~F.}\ \bibnamefont {Bell}}, \bibinfo {author} {\bibfnamefont {E.}~\bibnamefont {Pierpaoli}},\ and\ \bibinfo {author} {\bibfnamefont {K.}~\bibnamefont {Sigurdson}},\ }\href {https://doi.org/10.1103/PhysRevD.73.063523} {\bibfield  {journal} {\bibinfo  {journal} {Phys. Rev. D}\ }\textbf {\bibinfo {volume} {73}},\ \bibinfo {pages} {063523} (\bibinfo {year} {2006})}\BibitemShut {NoStop}%
\bibitem [{\citenamefont {Archidiacono}\ and\ \citenamefont {Hannestad}(2014)}]{Archidiacono2014}%
  \BibitemOpen
  \bibfield  {author} {\bibinfo {author} {\bibfnamefont {M.}~\bibnamefont {Archidiacono}}\ and\ \bibinfo {author} {\bibfnamefont {S.}~\bibnamefont {Hannestad}},\ }\href {https://doi.org/10.1088/1475-7516/2014/07/046} {\bibfield  {journal} {\bibinfo  {journal} {Journal of Cosmology and Astroparticle Physics}\ }\textbf {\bibinfo {volume} {2014}}\bibinfo  {number} { (07)},\ \bibinfo {pages} {046}}\BibitemShut {NoStop}%
\bibitem [{\citenamefont {Cyr-Racine}\ and\ \citenamefont {Sigurdson}(2014)}]{CyrRacine2014}%
  \BibitemOpen
\bibfield  {number} {  }\bibfield  {author} {\bibinfo {author} {\bibfnamefont {F.-Y.}\ \bibnamefont {Cyr-Racine}}\ and\ \bibinfo {author} {\bibfnamefont {K.}~\bibnamefont {Sigurdson}},\ }\href {https://doi.org/10.1103/PhysRevD.90.123533} {\bibfield  {journal} {\bibinfo  {journal} {Phys. Rev. D}\ }\textbf {\bibinfo {volume} {90}},\ \bibinfo {pages} {123533} (\bibinfo {year} {2014})}\BibitemShut {NoStop}%
\bibitem [{\citenamefont {Lancaster}\ \emph {et~al.}(2017{\natexlab{a}})\citenamefont {Lancaster}, \citenamefont {Cyr-Racine}, \citenamefont {Knox},\ and\ \citenamefont {Pan}}]{Lancaster_2017}%
  \BibitemOpen
  \bibfield  {author} {\bibinfo {author} {\bibfnamefont {L.}~\bibnamefont {Lancaster}}, \bibinfo {author} {\bibfnamefont {F.-Y.}\ \bibnamefont {Cyr-Racine}}, \bibinfo {author} {\bibfnamefont {L.}~\bibnamefont {Knox}},\ and\ \bibinfo {author} {\bibfnamefont {Z.}~\bibnamefont {Pan}},\ }\href {https://doi.org/10.1088/1475-7516/2017/07/033} {\bibfield  {journal} {\bibinfo  {journal} {Journal of Cosmology and Astroparticle Physics}\ }\textbf {\bibinfo {volume} {2017}}\bibinfo  {number} { (07)},\ \bibinfo {pages} {033}}\BibitemShut {NoStop}%
\bibitem [{\citenamefont {Park}\ \emph {et~al.}(2019)\citenamefont {Park}, \citenamefont {Kreisch}, \citenamefont {Dunkley}, \citenamefont {Hadzhiyska},\ and\ \citenamefont {Cyr-Racine}}]{Park2019}%
  \BibitemOpen
\bibfield  {number} {  }\bibfield  {author} {\bibinfo {author} {\bibfnamefont {M.}~\bibnamefont {Park}}, \bibinfo {author} {\bibfnamefont {C.~D.}\ \bibnamefont {Kreisch}}, \bibinfo {author} {\bibfnamefont {J.}~\bibnamefont {Dunkley}}, \bibinfo {author} {\bibfnamefont {B.}~\bibnamefont {Hadzhiyska}},\ and\ \bibinfo {author} {\bibfnamefont {F.-Y.}\ \bibnamefont {Cyr-Racine}},\ }\href {https://doi.org/10.1103/PhysRevD.100.063524} {\bibfield  {journal} {\bibinfo  {journal} {Phys. Rev. D}\ }\textbf {\bibinfo {volume} {100}},\ \bibinfo {pages} {063524} (\bibinfo {year} {2019})}\BibitemShut {NoStop}%
\bibitem [{\citenamefont {Escudero}\ and\ \citenamefont {Witte}(2020)}]{Escudero:2019gvw}%
  \BibitemOpen
  \bibfield  {author} {\bibinfo {author} {\bibfnamefont {M.}~\bibnamefont {Escudero}}\ and\ \bibinfo {author} {\bibfnamefont {S.~J.}\ \bibnamefont {Witte}},\ }\href {https://doi.org/10.1140/epjc/s10052-020-7854-5} {\bibfield  {journal} {\bibinfo  {journal} {Eur. Phys. J. C}\ }\textbf {\bibinfo {volume} {80}},\ \bibinfo {pages} {294} (\bibinfo {year} {2020})},\ \Eprint {https://arxiv.org/abs/1909.04044} {arXiv:1909.04044 [astro-ph.CO]} \BibitemShut {NoStop}%
\bibitem [{\citenamefont {Kreisch}\ \emph {et~al.}(2020)\citenamefont {Kreisch}, \citenamefont {Cyr-Racine},\ and\ \citenamefont {Dor\'e}}]{Kreisch2020}%
  \BibitemOpen
  \bibfield  {author} {\bibinfo {author} {\bibfnamefont {C.~D.}\ \bibnamefont {Kreisch}}, \bibinfo {author} {\bibfnamefont {F.-Y.}\ \bibnamefont {Cyr-Racine}},\ and\ \bibinfo {author} {\bibfnamefont {O.}~\bibnamefont {Dor\'e}},\ }\href {https://doi.org/10.1103/PhysRevD.101.123505} {\bibfield  {journal} {\bibinfo  {journal} {Phys. Rev. D}\ }\textbf {\bibinfo {volume} {101}},\ \bibinfo {pages} {123505} (\bibinfo {year} {2020})}\BibitemShut {NoStop}%
\bibitem [{\citenamefont {Knox}\ and\ \citenamefont {Millea}(2020)}]{knox2020hubble}%
  \BibitemOpen
  \bibfield  {author} {\bibinfo {author} {\bibfnamefont {L.}~\bibnamefont {Knox}}\ and\ \bibinfo {author} {\bibfnamefont {M.}~\bibnamefont {Millea}},\ }\href@noop {} {\bibfield  {journal} {\bibinfo  {journal} {Physical Review D}\ }\textbf {\bibinfo {volume} {101}},\ \bibinfo {pages} {043533} (\bibinfo {year} {2020})}\BibitemShut {NoStop}%
\bibitem [{\citenamefont {Choudhury}\ \emph {et~al.}(2021)\citenamefont {Choudhury}, \citenamefont {Hannestad},\ and\ \citenamefont {Tram}}]{Choudhury2021}%
  \BibitemOpen
  \bibfield  {author} {\bibinfo {author} {\bibfnamefont {S.~R.}\ \bibnamefont {Choudhury}}, \bibinfo {author} {\bibfnamefont {S.}~\bibnamefont {Hannestad}},\ and\ \bibinfo {author} {\bibfnamefont {T.}~\bibnamefont {Tram}},\ }\href {https://doi.org/10.1088/1475-7516/2021/03/084} {\bibfield  {journal} {\bibinfo  {journal} {Journal of Cosmology and Astroparticle Physics}\ }\textbf {\bibinfo {volume} {2021}}\bibinfo  {number} { (03)},\ \bibinfo {pages} {084}}\BibitemShut {NoStop}%
\bibitem [{\citenamefont {Venzor}\ \emph {et~al.}(2023)\citenamefont {Venzor}, \citenamefont {Garcia-Arroyo}, \citenamefont {De-Santiago},\ and\ \citenamefont {P{\'e}rez-Lorenzana}}]{Venzor:2023aka}%
  \BibitemOpen
\bibfield  {number} {  }\bibfield  {author} {\bibinfo {author} {\bibfnamefont {J.}~\bibnamefont {Venzor}}, \bibinfo {author} {\bibfnamefont {G.}~\bibnamefont {Garcia-Arroyo}}, \bibinfo {author} {\bibfnamefont {J.}~\bibnamefont {De-Santiago}},\ and\ \bibinfo {author} {\bibfnamefont {A.}~\bibnamefont {P{\'e}rez-Lorenzana}},\ }\href {https://doi.org/10.1103/PhysRevD.108.043536} {\bibfield  {journal} {\bibinfo  {journal} {Phys. Rev. D}\ }\textbf {\bibinfo {volume} {108}},\ \bibinfo {pages} {043536} (\bibinfo {year} {2023})},\ \Eprint {https://arxiv.org/abs/2303.12792} {arXiv:2303.12792 [astro-ph.CO]} \BibitemShut {NoStop}%
\bibitem [{\citenamefont {Oldengott}\ \emph {et~al.}(2015)\citenamefont {Oldengott}, \citenamefont {Rampf},\ and\ \citenamefont {Wong}}]{Oldengott2015}%
  \BibitemOpen
  \bibfield  {author} {\bibinfo {author} {\bibfnamefont {I.~M.}\ \bibnamefont {Oldengott}}, \bibinfo {author} {\bibfnamefont {C.}~\bibnamefont {Rampf}},\ and\ \bibinfo {author} {\bibfnamefont {Y.~Y.}\ \bibnamefont {Wong}},\ }\href {https://doi.org/10.1088/1475-7516/2015/04/016} {\bibfield  {journal} {\bibinfo  {journal} {Journal of Cosmology and Astroparticle Physics}\ }\textbf {\bibinfo {volume} {2015}}\bibinfo  {number} { (04)},\ \bibinfo {pages} {016}}\BibitemShut {NoStop}%
\bibitem [{\citenamefont {Lancaster}\ \emph {et~al.}(2017{\natexlab{b}})\citenamefont {Lancaster}, \citenamefont {Cyr-Racine}, \citenamefont {Knox},\ and\ \citenamefont {Pan}}]{Lancaster2017}%
  \BibitemOpen
\bibfield  {number} {  }\bibfield  {author} {\bibinfo {author} {\bibfnamefont {L.}~\bibnamefont {Lancaster}}, \bibinfo {author} {\bibfnamefont {F.-Y.}\ \bibnamefont {Cyr-Racine}}, \bibinfo {author} {\bibfnamefont {L.}~\bibnamefont {Knox}},\ and\ \bibinfo {author} {\bibfnamefont {Z.}~\bibnamefont {Pan}},\ }\href {https://doi.org/10.1088/1475-7516/2017/07/033} {\bibfield  {journal} {\bibinfo  {journal} {Journal of Cosmology and Astroparticle Physics}\ }\textbf {\bibinfo {volume} {2017}}\bibinfo  {number} { (07)},\ \bibinfo {pages} {033}}\BibitemShut {NoStop}%
\bibitem [{\citenamefont {Oldengott}\ \emph {et~al.}(2017)\citenamefont {Oldengott}, \citenamefont {Tram}, \citenamefont {Rampf},\ and\ \citenamefont {Wong}}]{Oldengott2017}%
  \BibitemOpen
\bibfield  {number} {  }\bibfield  {author} {\bibinfo {author} {\bibfnamefont {I.~M.}\ \bibnamefont {Oldengott}}, \bibinfo {author} {\bibfnamefont {T.}~\bibnamefont {Tram}}, \bibinfo {author} {\bibfnamefont {C.}~\bibnamefont {Rampf}},\ and\ \bibinfo {author} {\bibfnamefont {Y.~Y.}\ \bibnamefont {Wong}},\ }\href {https://doi.org/10.1088/1475-7516/2017/11/027} {\bibfield  {journal} {\bibinfo  {journal} {Journal of Cosmology and Astroparticle Physics}\ }\textbf {\bibinfo {volume} {2017}}\bibinfo  {number} { (11)},\ \bibinfo {pages} {027}}\BibitemShut {NoStop}%
\bibitem [{\citenamefont {Beacom}\ \emph {et~al.}(2004)\citenamefont {Beacom}, \citenamefont {Bell},\ and\ \citenamefont {Dodelson}}]{Beacom2004}%
  \BibitemOpen
\bibfield  {number} {  }\bibfield  {author} {\bibinfo {author} {\bibfnamefont {J.~F.}\ \bibnamefont {Beacom}}, \bibinfo {author} {\bibfnamefont {N.~F.}\ \bibnamefont {Bell}},\ and\ \bibinfo {author} {\bibfnamefont {S.}~\bibnamefont {Dodelson}},\ }\href {https://doi.org/10.1103/PhysRevLett.93.121302} {\bibfield  {journal} {\bibinfo  {journal} {Phys. Rev. Lett.}\ }\textbf {\bibinfo {volume} {93}},\ \bibinfo {pages} {121302} (\bibinfo {year} {2004})}\BibitemShut {NoStop}%
\bibitem [{\citenamefont {Hannestad}(2005)}]{Hannestad2005}%
  \BibitemOpen
  \bibfield  {author} {\bibinfo {author} {\bibfnamefont {S.}~\bibnamefont {Hannestad}},\ }\href {https://doi.org/10.1088/1475-7516/2005/02/011} {\bibfield  {journal} {\bibinfo  {journal} {Journal of Cosmology and Astroparticle Physics}\ }\textbf {\bibinfo {volume} {2005}}\bibinfo  {number} { (02)},\ \bibinfo {pages} {011}}\BibitemShut {NoStop}%
\bibitem [{\citenamefont {Chacko}\ \emph {et~al.}(2020)\citenamefont {Chacko}, \citenamefont {Dev}, \citenamefont {Du}, \citenamefont {Poulin},\ and\ \citenamefont {Tsai}}]{Chacko2020}%
  \BibitemOpen
\bibfield  {number} {  }\bibfield  {author} {\bibinfo {author} {\bibfnamefont {Z.}~\bibnamefont {Chacko}}, \bibinfo {author} {\bibfnamefont {A.}~\bibnamefont {Dev}}, \bibinfo {author} {\bibfnamefont {P.}~\bibnamefont {Du}}, \bibinfo {author} {\bibfnamefont {V.}~\bibnamefont {Poulin}},\ and\ \bibinfo {author} {\bibfnamefont {Y.}~\bibnamefont {Tsai}},\ }\href {https://doi.org/10.1007/JHEP04(2020)020} {\bibfield  {journal} {\bibinfo  {journal} {Journal of High Energy Physics}\ }\textbf {\bibinfo {volume} {2020}},\ \bibinfo {pages} {20} (\bibinfo {year} {2020})}\BibitemShut {NoStop}%
\bibitem [{\citenamefont {Escudero}\ \emph {et~al.}(2020)\citenamefont {Escudero}, \citenamefont {Lopez-Pavon}, \citenamefont {Rius},\ and\ \citenamefont {Sandner}}]{EscuderoRelaxing}%
  \BibitemOpen
  \bibfield  {author} {\bibinfo {author} {\bibfnamefont {M.}~\bibnamefont {Escudero}}, \bibinfo {author} {\bibfnamefont {J.}~\bibnamefont {Lopez-Pavon}}, \bibinfo {author} {\bibfnamefont {N.}~\bibnamefont {Rius}},\ and\ \bibinfo {author} {\bibfnamefont {S.}~\bibnamefont {Sandner}},\ }\href {https://doi.org/10.1007/JHEP12(2020)119} {\bibfield  {journal} {\bibinfo  {journal} {Journal of High Energy Physics}\ }\textbf {\bibinfo {volume} {2020}},\ \bibinfo {pages} {119} (\bibinfo {year} {2020})}\BibitemShut {NoStop}%
\bibitem [{\citenamefont {Esteban}\ and\ \citenamefont {Salvado}(2021)}]{Esteban2021}%
  \BibitemOpen
  \bibfield  {author} {\bibinfo {author} {\bibfnamefont {I.}~\bibnamefont {Esteban}}\ and\ \bibinfo {author} {\bibfnamefont {J.}~\bibnamefont {Salvado}},\ }\href {https://doi.org/10.1088/1475-7516/2021/05/036} {\bibfield  {journal} {\bibinfo  {journal} {Journal of Cosmology and Astroparticle Physics}\ }\textbf {\bibinfo {volume} {2021}}\bibinfo  {number} { (05)},\ \bibinfo {pages} {036}}\BibitemShut {NoStop}%
\bibitem [{\citenamefont {Barenboim}\ \emph {et~al.}(2021)\citenamefont {Barenboim}, \citenamefont {Chen}, \citenamefont {Hannestad}, \citenamefont {Oldengott}, \citenamefont {Tram},\ and\ \citenamefont {Wong}}]{Barenboim2021}%
  \BibitemOpen
\bibfield  {number} {  }\bibfield  {author} {\bibinfo {author} {\bibfnamefont {G.}~\bibnamefont {Barenboim}}, \bibinfo {author} {\bibfnamefont {J.~Z.}\ \bibnamefont {Chen}}, \bibinfo {author} {\bibfnamefont {S.}~\bibnamefont {Hannestad}}, \bibinfo {author} {\bibfnamefont {I.~M.}\ \bibnamefont {Oldengott}}, \bibinfo {author} {\bibfnamefont {T.}~\bibnamefont {Tram}},\ and\ \bibinfo {author} {\bibfnamefont {Y.~Y.}\ \bibnamefont {Wong}},\ }\href {https://doi.org/10.1088/1475-7516/2021/03/087} {\bibfield  {journal} {\bibinfo  {journal} {Journal of Cosmology and Astroparticle Physics}\ }\textbf {\bibinfo {volume} {2021}}\bibinfo  {number} { (03)},\ \bibinfo {pages} {087}}\BibitemShut {NoStop}%
\bibitem [{\citenamefont {Abell{\'a}n}\ \emph {et~al.}(2021)\citenamefont {Abell{\'a}n}, \citenamefont {Chacko}, \citenamefont {Dev}, \citenamefont {Du}, \citenamefont {Poulin},\ and\ \citenamefont {Tsai}}]{abellan2021}%
  \BibitemOpen
\bibfield  {number} {  }\bibfield  {author} {\bibinfo {author} {\bibfnamefont {G.~F.}\ \bibnamefont {Abell{\'a}n}}, \bibinfo {author} {\bibfnamefont {Z.}~\bibnamefont {Chacko}}, \bibinfo {author} {\bibfnamefont {A.}~\bibnamefont {Dev}}, \bibinfo {author} {\bibfnamefont {P.}~\bibnamefont {Du}}, \bibinfo {author} {\bibfnamefont {V.}~\bibnamefont {Poulin}},\ and\ \bibinfo {author} {\bibfnamefont {Y.}~\bibnamefont {Tsai}},\ }\href@noop {} {\bibfield  {journal} {\bibinfo  {journal} {arXiv preprint arXiv:2112.13862}\ } (\bibinfo {year} {2021})}\BibitemShut {NoStop}%
\bibitem [{\citenamefont {Franco~Abell{\'a}n}(2026)}]{FrancoAbellan:2026ori}%
  \BibitemOpen
  \bibfield  {author} {\bibinfo {author} {\bibfnamefont {G.}~\bibnamefont {Franco~Abell{\'a}n}},\ }\href {https://doi.org/10.1103/lwdv-qrcc} {\bibfield  {journal} {\bibinfo  {journal} {Phys. Rev. D}\ }\textbf {\bibinfo {volume} {113}},\ \bibinfo {pages} {123527} (\bibinfo {year} {2026})},\ \Eprint {https://arxiv.org/abs/2601.04312} {arXiv:2601.04312 [astro-ph.CO]} \BibitemShut {NoStop}%
\bibitem [{\citenamefont {Noriega}\ \emph {et~al.}(2025)\citenamefont {Noriega}, \citenamefont {De-Santiago}, \citenamefont {Garcia-Arroyo}, \citenamefont {Venzor},\ and\ \citenamefont {P{\'e}rez-Lorenzana}}]{Noriega:2025ulc}%
  \BibitemOpen
  \bibfield  {author} {\bibinfo {author} {\bibfnamefont {H.~E.}\ \bibnamefont {Noriega}}, \bibinfo {author} {\bibfnamefont {J.}~\bibnamefont {De-Santiago}}, \bibinfo {author} {\bibfnamefont {G.}~\bibnamefont {Garcia-Arroyo}}, \bibinfo {author} {\bibfnamefont {J.}~\bibnamefont {Venzor}},\ and\ \bibinfo {author} {\bibfnamefont {A.}~\bibnamefont {P{\'e}rez-Lorenzana}},\ }\href {https://doi.org/10.1103/b9x4-hnqn} {\bibfield  {journal} {\bibinfo  {journal} {Phys. Rev. D}\ }\textbf {\bibinfo {volume} {112}},\ \bibinfo {pages} {063509} (\bibinfo {year} {2025})},\ \Eprint {https://arxiv.org/abs/2506.07994} {arXiv:2506.07994 [astro-ph.CO]} \BibitemShut {NoStop}%
\bibitem [{\citenamefont {Perez-Castro}\ \emph {et~al.}(2026)\citenamefont {Perez-Castro}, \citenamefont {De-Santiago}, \citenamefont {Garcia-Arroyo}, \citenamefont {Venzor},\ and\ \citenamefont {Perez-Lorenzana}}]{Perez-Castro:2026muj}%
  \BibitemOpen
  \bibfield  {author} {\bibinfo {author} {\bibfnamefont {I.}~\bibnamefont {Perez-Castro}}, \bibinfo {author} {\bibfnamefont {J.}~\bibnamefont {De-Santiago}}, \bibinfo {author} {\bibfnamefont {G.}~\bibnamefont {Garcia-Arroyo}}, \bibinfo {author} {\bibfnamefont {J.}~\bibnamefont {Venzor}},\ and\ \bibinfo {author} {\bibfnamefont {A.}~\bibnamefont {Perez-Lorenzana}},\ }\Eprint {https://arxiv.org/abs/2602.12477} {arXiv:2602.12477 [hep-ph]}  (\bibinfo {year} {2026}),\ \bibinfo {note} {preprint}\BibitemShut {NoStop}%
\bibitem [{\citenamefont {Forastieri}\ \emph {et~al.}(2015)\citenamefont {Forastieri}, \citenamefont {Lattanzi},\ and\ \citenamefont {Natoli}}]{Forastieri2015}%
  \BibitemOpen
  \bibfield  {author} {\bibinfo {author} {\bibfnamefont {F.}~\bibnamefont {Forastieri}}, \bibinfo {author} {\bibfnamefont {M.}~\bibnamefont {Lattanzi}},\ and\ \bibinfo {author} {\bibfnamefont {P.}~\bibnamefont {Natoli}},\ }\href {https://doi.org/10.1088/1475-7516/2015/07/014} {\bibfield  {journal} {\bibinfo  {journal} {Journal of Cosmology and Astroparticle Physics}\ }\textbf {\bibinfo {volume} {2015}}\bibinfo  {number} { (07)},\ \bibinfo {pages} {014}}\BibitemShut {NoStop}%
\bibitem [{\citenamefont {Forastieri}\ \emph {et~al.}(2019)\citenamefont {Forastieri}, \citenamefont {Lattanzi},\ and\ \citenamefont {Natoli}}]{Forastieri:2019cuf}%
  \BibitemOpen
\bibfield  {number} {  }\bibfield  {author} {\bibinfo {author} {\bibfnamefont {F.}~\bibnamefont {Forastieri}}, \bibinfo {author} {\bibfnamefont {M.}~\bibnamefont {Lattanzi}},\ and\ \bibinfo {author} {\bibfnamefont {P.}~\bibnamefont {Natoli}},\ }\href {https://doi.org/10.1103/PhysRevD.100.103526} {\bibfield  {journal} {\bibinfo  {journal} {Phys. Rev. D}\ }\textbf {\bibinfo {volume} {100}},\ \bibinfo {pages} {103526} (\bibinfo {year} {2019})},\ \Eprint {https://arxiv.org/abs/1904.07810} {arXiv:1904.07810 [astro-ph.CO]} \BibitemShut {NoStop}%
\bibitem [{\citenamefont {Venzor}\ \emph {et~al.}(2022)\citenamefont {Venzor}, \citenamefont {Garcia-Arroyo}, \citenamefont {P{\'e}rez-Lorenzana},\ and\ \citenamefont {De-Santiago}}]{Venzor:2022hql}%
  \BibitemOpen
  \bibfield  {author} {\bibinfo {author} {\bibfnamefont {J.}~\bibnamefont {Venzor}}, \bibinfo {author} {\bibfnamefont {G.}~\bibnamefont {Garcia-Arroyo}}, \bibinfo {author} {\bibfnamefont {A.}~\bibnamefont {P{\'e}rez-Lorenzana}},\ and\ \bibinfo {author} {\bibfnamefont {J.}~\bibnamefont {De-Santiago}},\ }\href {https://doi.org/10.1103/PhysRevD.105.123539} {\bibfield  {journal} {\bibinfo  {journal} {Phys. Rev. D}\ }\textbf {\bibinfo {volume} {105}},\ \bibinfo {pages} {123539} (\bibinfo {year} {2022})},\ \Eprint {https://arxiv.org/abs/2202.09310} {arXiv:2202.09310 [astro-ph.CO]} \BibitemShut {NoStop}%
\bibitem [{\citenamefont {Venzor}\ \emph {et~al.}(2021)\citenamefont {Venzor}, \citenamefont {P{\'e}rez-Lorenzana},\ and\ \citenamefont {De-Santiago}}]{Venzor:2020ova}%
  \BibitemOpen
  \bibfield  {author} {\bibinfo {author} {\bibfnamefont {J.}~\bibnamefont {Venzor}}, \bibinfo {author} {\bibfnamefont {A.}~\bibnamefont {P{\'e}rez-Lorenzana}},\ and\ \bibinfo {author} {\bibfnamefont {J.}~\bibnamefont {De-Santiago}},\ }\href {https://doi.org/10.1103/PhysRevD.103.043534} {\bibfield  {journal} {\bibinfo  {journal} {Phys. Rev. D}\ }\textbf {\bibinfo {volume} {103}},\ \bibinfo {pages} {43534} (\bibinfo {year} {2021})},\ \Eprint {https://arxiv.org/abs/2009.08104} {arXiv:2009.08104 [hep-ph]} \BibitemShut {NoStop}%
\bibitem [{\citenamefont {Ge}\ and\ \citenamefont {Parke}(2019)}]{Ge2019}%
  \BibitemOpen
  \bibfield  {author} {\bibinfo {author} {\bibfnamefont {S.-F.}\ \bibnamefont {Ge}}\ and\ \bibinfo {author} {\bibfnamefont {S.~J.}\ \bibnamefont {Parke}},\ }\href {https://doi.org/10.1103/PhysRevLett.122.211801} {\bibfield  {journal} {\bibinfo  {journal} {Phys. Rev. Lett.}\ }\textbf {\bibinfo {volume} {122}},\ \bibinfo {pages} {211801} (\bibinfo {year} {2019})}\BibitemShut {NoStop}%
\bibitem [{\citenamefont {Babu}\ \emph {et~al.}(2020)\citenamefont {Babu}, \citenamefont {Chauhan},\ and\ \citenamefont {Dev}}]{Babu2020scalar}%
  \BibitemOpen
  \bibfield  {author} {\bibinfo {author} {\bibfnamefont {K.~S.}\ \bibnamefont {Babu}}, \bibinfo {author} {\bibfnamefont {G.}~\bibnamefont {Chauhan}},\ and\ \bibinfo {author} {\bibfnamefont {P.~S.~B.}\ \bibnamefont {Dev}},\ }\href {https://doi.org/10.1103/PhysRevD.101.095029} {\bibfield  {journal} {\bibinfo  {journal} {Phys. Rev. D}\ }\textbf {\bibinfo {volume} {101}},\ \bibinfo {pages} {095029} (\bibinfo {year} {2020})}\BibitemShut {NoStop}%
\bibitem [{\citenamefont {Verma}\ \emph {et~al.}(2026)\citenamefont {Verma}, \citenamefont {Arg{\"u}elles}, \citenamefont {Dev}, \citenamefont {Dutta},\ and\ \citenamefont {Martinez-Soler}}]{Verma:2026wrs}%
  \BibitemOpen
  \bibfield  {author} {\bibinfo {author} {\bibfnamefont {A.}~\bibnamefont {Verma}}, \bibinfo {author} {\bibfnamefont {C.~A.}\ \bibnamefont {Arg{\"u}elles}}, \bibinfo {author} {\bibfnamefont {P.~S.~B.}\ \bibnamefont {Dev}}, \bibinfo {author} {\bibfnamefont {B.}~\bibnamefont {Dutta}},\ and\ \bibinfo {author} {\bibfnamefont {I.}~\bibnamefont {Martinez-Soler}},\ }\Eprint {https://arxiv.org/abs/2606.25050} {arXiv:2606.25050 [hep-ph]}  (\bibinfo {year} {2026}),\ \bibinfo {note} {arXiv:2606.25050 [hep-ph]}\BibitemShut {NoStop}%
\bibitem [{\citenamefont {Einstein}(1919)}]{Einstein:1919gv}%
  \BibitemOpen
  \bibfield  {author} {\bibinfo {author} {\bibfnamefont {A.}~\bibnamefont {Einstein}},\ }\href@noop {} {\bibfield  {journal} {\bibinfo  {journal} {Sitzungsber. Preuss. Akad. Wiss. Berlin (Math. Phys. )}\ }\textbf {\bibinfo {volume} {1919}},\ \bibinfo {pages} {349} (\bibinfo {year} {1919})}\BibitemShut {NoStop}%
\bibitem [{\citenamefont {Ellis}\ \emph {et~al.}(2011)\citenamefont {Ellis}, \citenamefont {Van~Elst}, \citenamefont {Murugan},\ and\ \citenamefont {Uzan}}]{ellis2011trace}%
  \BibitemOpen
  \bibfield  {author} {\bibinfo {author} {\bibfnamefont {G.~F.}\ \bibnamefont {Ellis}}, \bibinfo {author} {\bibfnamefont {H.}~\bibnamefont {Van~Elst}}, \bibinfo {author} {\bibfnamefont {J.}~\bibnamefont {Murugan}},\ and\ \bibinfo {author} {\bibfnamefont {J.-P.}\ \bibnamefont {Uzan}},\ }\href@noop {} {\bibfield  {journal} {\bibinfo  {journal} {Classical and Quantum Gravity}\ }\textbf {\bibinfo {volume} {28}},\ \bibinfo {pages} {225007} (\bibinfo {year} {2011})}\BibitemShut {NoStop}%
\bibitem [{\citenamefont {Ellis}(2014)}]{Ellis:2013uxa}%
  \BibitemOpen
  \bibfield  {author} {\bibinfo {author} {\bibfnamefont {G.~F.~R.}\ \bibnamefont {Ellis}},\ }\href {https://doi.org/10.1007/s10714-013-1619-5} {\bibfield  {journal} {\bibinfo  {journal} {Gen. Rel. Grav.}\ }\textbf {\bibinfo {volume} {46}},\ \bibinfo {pages} {1619} (\bibinfo {year} {2014})},\ \Eprint {https://arxiv.org/abs/1306.3021} {arXiv:1306.3021 [gr-qc]} \BibitemShut {NoStop}%
\bibitem [{\citenamefont {Josset}\ \emph {et~al.}(2017)\citenamefont {Josset}, \citenamefont {Perez},\ and\ \citenamefont {Sudarsky}}]{Josset:2016vrq}%
  \BibitemOpen
  \bibfield  {author} {\bibinfo {author} {\bibfnamefont {T.}~\bibnamefont {Josset}}, \bibinfo {author} {\bibfnamefont {A.}~\bibnamefont {Perez}},\ and\ \bibinfo {author} {\bibfnamefont {D.}~\bibnamefont {Sudarsky}},\ }\href {https://doi.org/10.1103/PhysRevLett.118.021102} {\bibfield  {journal} {\bibinfo  {journal} {Phys. Rev. Lett.}\ }\textbf {\bibinfo {volume} {118}},\ \bibinfo {pages} {21102} (\bibinfo {year} {2017})},\ \Eprint {https://arxiv.org/abs/1604.04183} {arXiv:1604.04183 [gr-qc]} \BibitemShut {NoStop}%
\bibitem [{\citenamefont {Perez}\ and\ \citenamefont {Sudarsky}(2019)}]{Perez:2017krv}%
  \BibitemOpen
  \bibfield  {author} {\bibinfo {author} {\bibfnamefont {A.}~\bibnamefont {Perez}}\ and\ \bibinfo {author} {\bibfnamefont {D.}~\bibnamefont {Sudarsky}},\ }\href {https://doi.org/10.1103/PhysRevLett.122.221302} {\bibfield  {journal} {\bibinfo  {journal} {Phys. Rev. Lett.}\ }\textbf {\bibinfo {volume} {122}},\ \bibinfo {pages} {221302} (\bibinfo {year} {2019})},\ \Eprint {https://arxiv.org/abs/1711.05183} {arXiv:1711.05183 [gr-qc]} \BibitemShut {NoStop}%
\bibitem [{\citenamefont {Perez}\ \emph {et~al.}(2018)\citenamefont {Perez}, \citenamefont {Sudarsky},\ and\ \citenamefont {Bjorken}}]{Perez:2018wlo}%
  \BibitemOpen
  \bibfield  {author} {\bibinfo {author} {\bibfnamefont {A.}~\bibnamefont {Perez}}, \bibinfo {author} {\bibfnamefont {D.}~\bibnamefont {Sudarsky}},\ and\ \bibinfo {author} {\bibfnamefont {J.~D.}\ \bibnamefont {Bjorken}},\ }\href {https://doi.org/10.1142/S0218271818460021} {\bibfield  {journal} {\bibinfo  {journal} {Int. J. Mod. Phys. D}\ }\textbf {\bibinfo {volume} {27}},\ \bibinfo {pages} {1846002} (\bibinfo {year} {2018})},\ \Eprint {https://arxiv.org/abs/1804.07162} {arXiv:1804.07162 [gr-qc]} \BibitemShut {NoStop}%
\bibitem [{\citenamefont {Corral}\ \emph {et~al.}(2020)\citenamefont {Corral}, \citenamefont {Cruz},\ and\ \citenamefont {Gonz{\'a}lez}}]{corral2020diffusion}%
  \BibitemOpen
  \bibfield  {author} {\bibinfo {author} {\bibfnamefont {C.}~\bibnamefont {Corral}}, \bibinfo {author} {\bibfnamefont {N.}~\bibnamefont {Cruz}},\ and\ \bibinfo {author} {\bibfnamefont {E.}~\bibnamefont {Gonz{\'a}lez}},\ }\href@noop {} {\bibfield  {journal} {\bibinfo  {journal} {Physical Review D}\ }\textbf {\bibinfo {volume} {102}},\ \bibinfo {pages} {023508} (\bibinfo {year} {2020})}\BibitemShut {NoStop}%
\bibitem [{\citenamefont {Linares~Cede{\~n}o}\ and\ \citenamefont {Nucamendi}(2021)}]{LinaresCedeno:2020uxx}%
  \BibitemOpen
  \bibfield  {author} {\bibinfo {author} {\bibfnamefont {F.~X.}\ \bibnamefont {Linares~Cede{\~n}o}}\ and\ \bibinfo {author} {\bibfnamefont {U.}~\bibnamefont {Nucamendi}},\ }\href {https://doi.org/10.1016/j.dark.2021.100807} {\bibfield  {journal} {\bibinfo  {journal} {Phys. Dark Univ.}\ }\textbf {\bibinfo {volume} {32}},\ \bibinfo {pages} {100807} (\bibinfo {year} {2021})},\ \Eprint {https://arxiv.org/abs/2009.10268} {arXiv:2009.10268 [astro-ph.CO]} \BibitemShut {NoStop}%
\bibitem [{\citenamefont {Landau}\ \emph {et~al.}(2023)\citenamefont {Landau}, \citenamefont {Benetti}, \citenamefont {Perez},\ and\ \citenamefont {Sudarsky}}]{Landau:2022mhm}%
  \BibitemOpen
  \bibfield  {author} {\bibinfo {author} {\bibfnamefont {S.~J.}\ \bibnamefont {Landau}}, \bibinfo {author} {\bibfnamefont {M.}~\bibnamefont {Benetti}}, \bibinfo {author} {\bibfnamefont {A.}~\bibnamefont {Perez}},\ and\ \bibinfo {author} {\bibfnamefont {D.}~\bibnamefont {Sudarsky}},\ }\href {https://doi.org/10.1103/PhysRevD.108.043524} {\bibfield  {journal} {\bibinfo  {journal} {Phys. Rev. D}\ }\textbf {\bibinfo {volume} {108}},\ \bibinfo {pages} {043524} (\bibinfo {year} {2023})},\ \Eprint {https://arxiv.org/abs/2211.07424} {arXiv:2211.07424 [astro-ph.CO]} \BibitemShut {NoStop}%
\bibitem [{\citenamefont {Weldon}(1982)}]{Weldon:1982bn}%
  \BibitemOpen
  \bibfield  {author} {\bibinfo {author} {\bibfnamefont {H.~A.}\ \bibnamefont {Weldon}},\ }\href {https://doi.org/10.1103/PhysRevD.26.2789} {\bibfield  {journal} {\bibinfo  {journal} {Phys. Rev. D}\ }\textbf {\bibinfo {volume} {26}},\ \bibinfo {pages} {2789} (\bibinfo {year} {1982})}\BibitemShut {NoStop}%
\bibitem [{\citenamefont {N{\"o}tzold}\ and\ \citenamefont {Raffelt}(1988)}]{NotzoldRaffelt1988}%
  \BibitemOpen
  \bibfield  {author} {\bibinfo {author} {\bibfnamefont {D.}~\bibnamefont {N{\"o}tzold}}\ and\ \bibinfo {author} {\bibfnamefont {G.}~\bibnamefont {Raffelt}},\ }\href {https://doi.org/10.1016/0550-3213(88)90113-7} {\bibfield  {journal} {\bibinfo  {journal} {Nucl. Phys. B}\ }\textbf {\bibinfo {volume} {307}},\ \bibinfo {pages} {924} (\bibinfo {year} {1988})}\BibitemShut {NoStop}%
\bibitem [{\citenamefont {Dodelson}\ and\ \citenamefont {Schmidt}(2025)}]{DodelsonSchmidt2025}%
  \BibitemOpen
  \bibfield  {author} {\bibinfo {author} {\bibfnamefont {S.}~\bibnamefont {Dodelson}}\ and\ \bibinfo {author} {\bibfnamefont {F.}~\bibnamefont {Schmidt}},\ }\href@noop {} {\emph {\bibinfo {title} {Modern Cosmology}}},\ \bibinfo {edition} {3rd}\ ed.\ (\bibinfo  {publisher} {Elsevier},\ \bibinfo {address} {London},\ \bibinfo {year} {2025})\BibitemShut {NoStop}%
\bibitem [{\citenamefont {Rezzolla}\ and\ \citenamefont {Zanotti}(2013)}]{Rezzolla:2013dea}%
  \BibitemOpen
  \bibfield  {author} {\bibinfo {author} {\bibfnamefont {L.}~\bibnamefont {Rezzolla}}\ and\ \bibinfo {author} {\bibfnamefont {O.}~\bibnamefont {Zanotti}},\ }\href {https://doi.org/10.1093/acprof:oso/9780198528906.001.0001} {\emph {\bibinfo {title} {{Relativistic Hydrodynamics}}}}\ (\bibinfo  {publisher} {Oxford University Press},\ \bibinfo {year} {2013})\BibitemShut {NoStop}%
\bibitem [{\citenamefont {Gariazzo}\ \emph {et~al.}(2018)\citenamefont {Gariazzo}, \citenamefont {Archidiacono}, \citenamefont {de~Salas}, \citenamefont {Mena}, \citenamefont {Ternes},\ and\ \citenamefont {T{\'o}rtola}}]{Gariazzo:2018pei}%
  \BibitemOpen
  \bibfield  {author} {\bibinfo {author} {\bibfnamefont {S.}~\bibnamefont {Gariazzo}}, \bibinfo {author} {\bibfnamefont {M.}~\bibnamefont {Archidiacono}}, \bibinfo {author} {\bibfnamefont {P.~F.}\ \bibnamefont {de~Salas}}, \bibinfo {author} {\bibfnamefont {O.}~\bibnamefont {Mena}}, \bibinfo {author} {\bibfnamefont {C.~A.}\ \bibnamefont {Ternes}},\ and\ \bibinfo {author} {\bibfnamefont {M.}~\bibnamefont {T{\'o}rtola}},\ }\href {https://doi.org/10.1088/1475-7516/2018/03/011} {\bibfield  {journal} {\bibinfo  {journal} {JCAP}\ }\textbf {\bibinfo {volume} {03}}\bibfield  {number} {\bibinfo  {number} { (03)},\ \bibinfo {pages} {011}},\ }\Eprint {https://arxiv.org/abs/1801.04946} {arXiv:1801.04946 [hep-ph]} \BibitemShut {NoStop}%
\bibitem [{\citenamefont {Long}\ \emph {et~al.}(2018)\citenamefont {Long}, \citenamefont {Raveri}, \citenamefont {Hu},\ and\ \citenamefont {Dodelson}}]{Long:2017dru}%
  \BibitemOpen
  \bibfield  {author} {\bibinfo {author} {\bibfnamefont {A.~J.}\ \bibnamefont {Long}}, \bibinfo {author} {\bibfnamefont {M.}~\bibnamefont {Raveri}}, \bibinfo {author} {\bibfnamefont {W.}~\bibnamefont {Hu}},\ and\ \bibinfo {author} {\bibfnamefont {S.}~\bibnamefont {Dodelson}},\ }\href {https://doi.org/10.1103/PhysRevD.97.043510} {\bibfield  {journal} {\bibinfo  {journal} {Phys. Rev. D}\ }\textbf {\bibinfo {volume} {97}},\ \bibinfo {pages} {043510} (\bibinfo {year} {2018})},\ \Eprint {https://arxiv.org/abs/1711.08434} {arXiv:1711.08434 [astro-ph.CO]} \BibitemShut {NoStop}%
\bibitem [{\citenamefont {Roy~Choudhury}\ and\ \citenamefont {Choubey}(2018)}]{RoyChoudhury:2018gay}%
  \BibitemOpen
  \bibfield  {author} {\bibinfo {author} {\bibfnamefont {S.}~\bibnamefont {Roy~Choudhury}}\ and\ \bibinfo {author} {\bibfnamefont {S.}~\bibnamefont {Choubey}},\ }\href {https://doi.org/10.1088/1475-7516/2018/09/017} {\bibfield  {journal} {\bibinfo  {journal} {JCAP}\ }\textbf {\bibinfo {volume} {09}}\bibfield  {number} {\bibinfo  {number} { (09)},\ \bibinfo {pages} {017}},\ }\Eprint {https://arxiv.org/abs/1806.10832} {arXiv:1806.10832 [astro-ph.CO]} \BibitemShut {NoStop}%
\bibitem [{\citenamefont {Hannestad}\ and\ \citenamefont {Schwetz}(2016)}]{Hannestad:2016fog}%
  \BibitemOpen
  \bibfield  {author} {\bibinfo {author} {\bibfnamefont {S.}~\bibnamefont {Hannestad}}\ and\ \bibinfo {author} {\bibfnamefont {T.}~\bibnamefont {Schwetz}},\ }\href {https://doi.org/10.1088/1475-7516/2016/11/035} {\bibfield  {journal} {\bibinfo  {journal} {JCAP}\ }\textbf {\bibinfo {volume} {11}}\bibfield  {number} {\bibinfo  {number} { (11)},\ \bibinfo {pages} {35}},\ }\Eprint {https://arxiv.org/abs/1606.04691} {arXiv:1606.04691 [astro-ph.CO]} \BibitemShut {NoStop}%
\bibitem [{\citenamefont {Loureiro}\ \emph {et~al.}(2019)\citenamefont {Loureiro}, \citenamefont {Cuceu}, \citenamefont {Abdalla}, \citenamefont {Moraes}, \citenamefont {Whiteway}, \citenamefont {McLeod}, \citenamefont {Balan}, \citenamefont {Lahav}, \citenamefont {Benoit-Levy}, \citenamefont {Manera}, \citenamefont {Rollins},\ and\ \citenamefont {Xavier}}]{Loureiro:2019hwh}%
  \BibitemOpen
  \bibfield  {author} {\bibinfo {author} {\bibfnamefont {A.}~\bibnamefont {Loureiro}}, \bibinfo {author} {\bibfnamefont {A.}~\bibnamefont {Cuceu}}, \bibinfo {author} {\bibfnamefont {F.~B.}\ \bibnamefont {Abdalla}}, \bibinfo {author} {\bibfnamefont {B.}~\bibnamefont {Moraes}}, \bibinfo {author} {\bibfnamefont {L.}~\bibnamefont {Whiteway}}, \bibinfo {author} {\bibfnamefont {M.}~\bibnamefont {McLeod}}, \bibinfo {author} {\bibfnamefont {S.~T.}\ \bibnamefont {Balan}}, \bibinfo {author} {\bibfnamefont {O.}~\bibnamefont {Lahav}}, \bibinfo {author} {\bibfnamefont {A.}~\bibnamefont {Benoit-Levy}}, \bibinfo {author} {\bibfnamefont {M.}~\bibnamefont {Manera}}, \bibinfo {author} {\bibfnamefont {R.~P.}\ \bibnamefont {Rollins}},\ and\ \bibinfo {author} {\bibfnamefont {H.~S.}\ \bibnamefont {Xavier}},\ }\href {https://doi.org/10.1103/PhysRevLett.123.081301} {\bibfield  {journal} {\bibinfo  {journal} {Phys. Rev. Lett.}\ }\textbf {\bibinfo {volume} {123}},\ \bibinfo {pages} {081301} (\bibinfo {year} {2019})}\BibitemShut
  {NoStop}%
\bibitem [{\citenamefont {Chebat}\ \emph {et~al.}(2026)\citenamefont {Chebat} \emph {et~al.}}]{Chebat:2025kes}%
  \BibitemOpen
  \bibfield  {author} {\bibinfo {author} {\bibfnamefont {D.}~\bibnamefont {Chebat}} \emph {et~al.},\ }\href {https://doi.org/10.1088/1475-7516/2026/01/041} {\bibfield  {journal} {\bibinfo  {journal} {JCAP}\ }\textbf {\bibinfo {volume} {01}}\bibfield  {number} {\bibinfo  {number} { (01)},\ \bibinfo {pages} {041}},\ }\Eprint {https://arxiv.org/abs/2507.12401} {arXiv:2507.12401 [astro-ph.CO]} \BibitemShut {NoStop}%
\bibitem [{\citenamefont {Pesce}\ \emph {et~al.}(2020)\citenamefont {Pesce}, \citenamefont {Braatz}, \citenamefont {Reid}, \citenamefont {Riess}, \citenamefont {Scolnic}, \citenamefont {Condon}, \citenamefont {Gao}, \citenamefont {Henkel}, \citenamefont {Impellizzeri}, \citenamefont {Kuo},\ and\ \citenamefont {Lo}}]{Pesce_2020_Megamasers}%
  \BibitemOpen
  \bibfield  {author} {\bibinfo {author} {\bibfnamefont {D.~W.}\ \bibnamefont {Pesce}}, \bibinfo {author} {\bibfnamefont {J.~A.}\ \bibnamefont {Braatz}}, \bibinfo {author} {\bibfnamefont {M.~J.}\ \bibnamefont {Reid}}, \bibinfo {author} {\bibfnamefont {A.~G.}\ \bibnamefont {Riess}}, \bibinfo {author} {\bibfnamefont {D.}~\bibnamefont {Scolnic}}, \bibinfo {author} {\bibfnamefont {J.~J.}\ \bibnamefont {Condon}}, \bibinfo {author} {\bibfnamefont {F.}~\bibnamefont {Gao}}, \bibinfo {author} {\bibfnamefont {C.}~\bibnamefont {Henkel}}, \bibinfo {author} {\bibfnamefont {C.~M.~V.}\ \bibnamefont {Impellizzeri}}, \bibinfo {author} {\bibfnamefont {C.~Y.}\ \bibnamefont {Kuo}},\ and\ \bibinfo {author} {\bibfnamefont {K.~Y.}\ \bibnamefont {Lo}},\ }\href {https://doi.org/10.3847/2041-8213/ab75f0} {\bibfield  {journal} {\bibinfo  {journal} {The Astrophysical Journal Letters}\ }\textbf {\bibinfo {volume} {891}},\ \bibinfo {pages} {L1} (\bibinfo {year} {2020})}\BibitemShut {NoStop}%
\bibitem [{\citenamefont {Moresco}\ \emph {et~al.}(2020{\natexlab{a}})\citenamefont {Moresco}, \citenamefont {Jimenez}, \citenamefont {Verde}, \citenamefont {Cimatti},\ and\ \citenamefont {Pozzetti}}]{Moresco:2020fbm}%
  \BibitemOpen
  \bibfield  {author} {\bibinfo {author} {\bibfnamefont {M.}~\bibnamefont {Moresco}}, \bibinfo {author} {\bibfnamefont {R.}~\bibnamefont {Jimenez}}, \bibinfo {author} {\bibfnamefont {L.}~\bibnamefont {Verde}}, \bibinfo {author} {\bibfnamefont {A.}~\bibnamefont {Cimatti}},\ and\ \bibinfo {author} {\bibfnamefont {L.}~\bibnamefont {Pozzetti}},\ }\href {https://doi.org/10.3847/1538-4357/ab9eb0} {\bibfield  {journal} {\bibinfo  {journal} {Astrophys. J.}\ }\textbf {\bibinfo {volume} {898}},\ \bibinfo {pages} {82} (\bibinfo {year} {2020}{\natexlab{a}})},\ \Eprint {https://arxiv.org/abs/2003.07362} {arXiv:2003.07362 [astro-ph.GA]} \BibitemShut {NoStop}%
\bibitem [{\citenamefont {Moresco}\ \emph {et~al.}(2020{\natexlab{b}})\citenamefont {Moresco}, \citenamefont {Jimenez}, \citenamefont {Verde}, \citenamefont {Cimatti},\ and\ \citenamefont {Pozzetti}}]{Moresco_Covariance}%
  \BibitemOpen
  \bibfield  {author} {\bibinfo {author} {\bibfnamefont {M.}~\bibnamefont {Moresco}}, \bibinfo {author} {\bibfnamefont {R.}~\bibnamefont {Jimenez}}, \bibinfo {author} {\bibfnamefont {L.}~\bibnamefont {Verde}}, \bibinfo {author} {\bibfnamefont {A.}~\bibnamefont {Cimatti}},\ and\ \bibinfo {author} {\bibfnamefont {L.}~\bibnamefont {Pozzetti}},\ }\href {https://doi.org/10.3847/1538-4357/ab9eb0} {\bibfield  {journal} {\bibinfo  {journal} {The Astrophysical Journal}\ }\textbf {\bibinfo {volume} {898}},\ \bibinfo {pages} {82} (\bibinfo {year} {2020}{\natexlab{b}})}\BibitemShut {NoStop}%
\bibitem [{\citenamefont {Popovic}\ \emph {et~al.}(2026)\citenamefont {Popovic}, \citenamefont {Shah}, \citenamefont {Kenworthy} \emph {et~al.}}]{10.1093/mnras/stag632_DES_Dovekie}%
  \BibitemOpen
  \bibfield  {author} {\bibinfo {author} {\bibfnamefont {B.}~\bibnamefont {Popovic}}, \bibinfo {author} {\bibfnamefont {P.}~\bibnamefont {Shah}}, \bibinfo {author} {\bibfnamefont {W.~D.}\ \bibnamefont {Kenworthy}}, \emph {et~al.} (\bibinfo {collaboration} {DES Collaboration}),\ }\bibfield  {journal} {\bibinfo  {journal} {Mon. Not. Roy. Astron. Soc.}\ }\textbf {\bibinfo {volume} {548}},\ \href {https://doi.org/10.1093/mnras/stag632} {10.1093/mnras/stag632} (\bibinfo {year} {2026})\BibitemShut {NoStop}%
\bibitem [{\citenamefont {Abbott}\ \emph {et~al.}(2024{\natexlab{b}})\citenamefont {Abbott}, \citenamefont {Acevedo}, \citenamefont {Aguena} \emph {et~al.}}]{Abbott_2024_DES_old}%
  \BibitemOpen
  \bibfield  {author} {\bibinfo {author} {\bibfnamefont {T.~M.~C.}\ \bibnamefont {Abbott}}, \bibinfo {author} {\bibfnamefont {M.}~\bibnamefont {Acevedo}}, \bibinfo {author} {\bibfnamefont {M.}~\bibnamefont {Aguena}}, \emph {et~al.} (\bibinfo {collaboration} {DES Collaboration}),\ }\href {https://doi.org/10.3847/2041-8213/ad6f9f} {\bibfield  {journal} {\bibinfo  {journal} {Astrophys. J. Lett.}\ }\textbf {\bibinfo {volume} {973}},\ \bibinfo {pages} {L14} (\bibinfo {year} {2024}{\natexlab{b}})}\BibitemShut {NoStop}%
\bibitem [{\citenamefont {Chen}\ \emph {et~al.}(2019)\citenamefont {Chen}, \citenamefont {Huang},\ and\ \citenamefont {Wang}}]{Chen_2019_Planck_DP}%
  \BibitemOpen
  \bibfield  {author} {\bibinfo {author} {\bibfnamefont {L.}~\bibnamefont {Chen}}, \bibinfo {author} {\bibfnamefont {Q.-G.}\ \bibnamefont {Huang}},\ and\ \bibinfo {author} {\bibfnamefont {K.}~\bibnamefont {Wang}},\ }\href {https://doi.org/10.1088/1475-7516/2019/02/028} {\bibfield  {journal} {\bibinfo  {journal} {Journal of Cosmology and Astroparticle Physics}\ }\textbf {\bibinfo {volume} {2019}}\bibinfo  {number} { (02)},\ \bibinfo {pages} {028}}\BibitemShut {NoStop}%
\bibitem [{Note1()}]{Note1}%
  \BibitemOpen
\bibfield  {number} {  }\bibinfo {note} {See \protect \url {https://github.com/MauLoHdz/LUDB}}\BibitemShut {NoStop}%
\bibitem [{\citenamefont {Workman}\ \emph {et~al.}(2022)\citenamefont {Workman} \emph {et~al.}}]{PDG2022}%
  \BibitemOpen
  \bibfield  {author} {\bibinfo {author} {\bibfnamefont {R.~L.}\ \bibnamefont {Workman}} \emph {et~al.},\ }\href@noop {} {\bibfield  {journal} {\bibinfo  {journal} {Progress of Theoretical and Experimental Physics}\ }\textbf {\bibinfo {volume} {2022}},\ \bibinfo {pages} {083C01} (\bibinfo {year} {2022})}\BibitemShut {NoStop}%
\bibitem [{\citenamefont {Cowan}(1998)}]{Cowan1998}%
  \BibitemOpen
  \bibfield  {author} {\bibinfo {author} {\bibfnamefont {G.}~\bibnamefont {Cowan}},\ }\href@noop {} {\emph {\bibinfo {title} {Statistical Data Analysis}}}\ (\bibinfo  {publisher} {Oxford University Press},\ \bibinfo {year} {1998})\BibitemShut {NoStop}%
\bibitem [{\citenamefont {Blinov}\ \emph {et~al.}(2019)\citenamefont {Blinov}, \citenamefont {Marques-Tavares},\ and\ \citenamefont {Perez-Gonzalez}}]{Blinov:2019gcj}%
  \BibitemOpen
  \bibfield  {author} {\bibinfo {author} {\bibfnamefont {N.}~\bibnamefont {Blinov}}, \bibinfo {author} {\bibfnamefont {G.}~\bibnamefont {Marques-Tavares}},\ and\ \bibinfo {author} {\bibfnamefont {Y.~F.}\ \bibnamefont {Perez-Gonzalez}},\ }\href@noop {} {\bibfield  {journal} {\bibinfo  {journal} {JCAP}\ }\textbf {\bibinfo {volume} {07}}\bibfield  {number} {\bibinfo  {number} { (07)},\ \bibinfo {pages} {029}},\ }\Eprint {https://arxiv.org/abs/1903.00047} {arXiv:1903.00047 [hep-ph]} \BibitemShut {NoStop}%
\bibitem [{\citenamefont {Berryman}\ \emph {et~al.}(2023)\citenamefont {Berryman} \emph {et~al.}}]{Berryman:2022hds}%
  \BibitemOpen
  \bibfield  {author} {\bibinfo {author} {\bibfnamefont {J.~M.}\ \bibnamefont {Berryman}} \emph {et~al.},\ }\href@noop {} {\bibfield  {journal} {\bibinfo  {journal} {SciPost Phys. Proc.}\ }\textbf {\bibinfo {volume} {12}},\ \bibinfo {pages} {042} (\bibinfo {year} {2023})},\ \Eprint {https://arxiv.org/abs/2203.01955} {arXiv:2203.01955 [hep-ph]} \BibitemShut {NoStop}%
\bibitem [{\citenamefont {O'Hare}(2024)}]{OHare:2024nmr}%
  \BibitemOpen
  \bibfield  {author} {\bibinfo {author} {\bibfnamefont {C.~A.~J.}\ \bibnamefont {O'Hare}},\ }\href {https://doi.org/10.22323/1.454.0040} {\bibfield  {journal} {\bibinfo  {journal} {PoS}\ }\textbf {\bibinfo {volume} {COSMICWISPers}},\ \bibinfo {pages} {040} (\bibinfo {year} {2024})},\ \Eprint {https://arxiv.org/abs/2403.17697} {arXiv:2403.17697 [hep-ph]} \BibitemShut {NoStop}%
\bibitem [{\citenamefont {Marsh}(2016)}]{Marsh2016}%
  \BibitemOpen
  \bibfield  {author} {\bibinfo {author} {\bibfnamefont {D.~J.~E.}\ \bibnamefont {Marsh}},\ }\href {https://doi.org/10.1016/j.physrep.2016.06.005} {\bibfield  {journal} {\bibinfo  {journal} {Phys. Rept.}\ }\textbf {\bibinfo {volume} {643}},\ \bibinfo {pages} {1} (\bibinfo {year} {2016})},\ \Eprint {https://arxiv.org/abs/1510.07633} {arXiv:1510.07633 [astro-ph.CO]} \BibitemShut {NoStop}%
\bibitem [{\citenamefont {Ferreira}(2021)}]{Ferreira2021}%
  \BibitemOpen
  \bibfield  {author} {\bibinfo {author} {\bibfnamefont {E.~G.~M.}\ \bibnamefont {Ferreira}},\ }\href {https://doi.org/10.1007/s00159-021-00135-6} {\bibfield  {journal} {\bibinfo  {journal} {Astron. Astrophys. Rev.}\ }\textbf {\bibinfo {volume} {29}},\ \bibinfo {pages} {7} (\bibinfo {year} {2021})},\ \Eprint {https://arxiv.org/abs/2005.03254} {arXiv:2005.03254 [astro-ph.CO]} \BibitemShut {NoStop}%
\end{thebibliography}%

\end{document}